\documentstyle[draft,epsf]{susspsty1.tex}%sussp
\begin{document}
%\begin{flushright}
%UCLA/11/TEP/94
%\end{flushright}
\headings{Physics Beyond the Standard Model$^*$}
%\footnote{Lectures given at the
%1993 Scottish Summer School, St. Andrews, Scotland, August 1993, and at the
%1993 Escuelas Latino Americano de F\'{\i}sica, Mar del Plata, Argentina, July
%1993.}}
{Beyond the Standard Model}
{Roberto Peccei}{Department of Physics, University of California Los Angeles \\
Los Angeles, California 90024-1547}
\vspace{1.5cm}

\begin{abstract}

In these lectures I describe some of the open questions in the standard model
relating to the nature and origin of mass, forces and matter and discuss some
of the speculative theoretical ideas put forth in this regard.  Some of the
topics touched upon include supersymmetry, dynamical symmetry breaking,
composite models, grand unified theories and superstrings.
\end{abstract}

\vspace{3.5cm}

{\scriptsize $^*$Lectures given at the 1993 Scottish Summer School, St.
Andrews,
Scotland, August 1993, and at the 1993 Escuela Latino Americano de
F\'{\i}sica, Mar del Plata, Argentina, July 1993.}
\newpage
%%%%%%%%%%%%%%%%%%%%%%%%%
%%%%%%%%%%%%%%% SECTION 1
%%%%%%%%%%%%%%%%%%%%%%%%%
%\input{partone.tex}
\section{Introduction:}

\subsection*{Central and Side Issues in the Standard Model}

The standard model ({\sc sm}),
the SU(3)$\times$SU(2)$\times$U(1) gauge
theory of the strong and electroweak interactions, has
proven to be extraordinarily robust. To date, all data are in
perfect accord with its predictions, with at best very indirect
hints for any physics beyond the {\sc sm}. This is well illustrated
by Fig.~1 where measurements of the weak mixing
angle done in experiments over an energy range from eV, to 100 GeV are
shown.  As can be seen all results are perfectly
consistent with each other.

\begin{figure*}[h]
\vspace{8cm}
\caption[]{Measured values of the weak mixing angle, from
\cite{Langacker}.}
\end{figure*}

Nevertheless, there are a number of important features
in the {\sc sm} which are poorly understood and some open
questions whose answers necessarily can only be found
by going beyond the {\sc sm}. To organize the discussion,
it seems useful to divide
the issues one wants answers to into two kinds: central
issues and side issues.

\subsection*{Central Issues}

These are questions whose answers are fundamental for our understanding
of the {\sc sm} and which, most likely, will require going beyond
the {\sc sm} for a proper answer. Broadly speaking there are three
such central questions:

\begin{enumerate}
\item The Question of Mass

\item The Question of Forces

\item The Question of Matter
\end{enumerate}

\noindent I will spend considerable time in what follows defining these
questions more clearly and trying to describe suggestive
answers to them.

\subsubsection*{Side Issues}

These are issues in the {\sc sm} which are either open or not
addressed, whose answer may involve extensions of the {\sc sm}
or, when better understood, may have an influence on the
deeper central issues. Some examples of questions in this
class are:
\begin{enumerate}
\item Do neutrinos have mass?
\item What is the dark matter in the universe?
\item What is the origin of the matter-antimatter
 asymmetry in the universe?
\item Why is there no strong CP violation?
\item Is the observed CP violation in the Kaon system only a result of
quark mixing?
\end{enumerate}
In these lectures, I will mostly focus on the central open questions in
the standard model, returning to discuss when appropriate how the
answers to these main questions may impinge on these side issues, and
vice versa.  I will also ignore, for the most part, open calculational
problems in the {\sc sm}.  Although problems like calculating the
spectrum of hadrons from QCD, or the matrix elements of weak operators
between hadronic states, are of great practical importance they are, in
some sense, not open problems.  Eventually, as our calculational tools
get further refined there is hope that these problems at least will be
resolved.
\section{The Question of Mass}

One of the fundamental attributes of both the elementary
excitations (quarks, leptons, and gauge bosons) and composite
states (protons, neutrons, pions,\ldots) present in the {\sc sm} is
that all of these particles have mass. However, the {\sc sm},
possesses only dimensionless coupling constants ($g_1$, $g_2$, $g_3$)
and no explicit mass parameters at all! How then are these masses
generated?

In natural units, $\hbar = c = 1$, the action $S$ is dimensionless.
 Since
\[
S = \int d^4x\,{\cal L}\ ,
\]
one sees that
the Lagrangian density ${\cal L}$ scales like (mass)$^4$. From the
kinetic energies of fermions

\[
{\cal L}_{\rm kinetic} =
-\overline{\psi}\gamma^\mu\sfrac{1}{i}\partial_{\mu}\psi
\]
and gauge fields
\[
{\cal L}_{\rm kinetic} = - \frac{1}{4}
\left(\partial^\mu A^\nu - \partial^\nu A^\mu\right) {\cdot}
\left(\partial_\mu A_\nu - \partial_\nu A_\mu\right) + {\cdots}
\]
one learns that $\psi \sim {\rm(mass)}^{3/2}$, while $A_{\mu} \sim {\rm
(mass)}^1$.
Interactions in the {\sc sm} are introduced via minimal substitution
\[
\partial_\mu \to \partial_\mu - i\,g_i A_{i \mu}
\]
and so the coupling constants, $g_i$, are dimensionless.
Obviously no mass scales enter in ${\cal L}_{\rm kinetic}$ or in
${\cal L}_{\rm interaction}$.
Furthermore, no explicit mass terms are allowed, both theoretically
and experimentally, in the standard model since:
\begin{description}
\item{i)} The gauge principle, $S \to S$ as
${\dst A^\mu \to A^\mu - \frac{i}{g}\partial^\mu\Lambda}$,
forbids explicit mass terms for the
SU(3)$\times$SU(2)$\times$U(1) gauge bosons.

\item{ii)} The observed (V$-$A) nature of the weak interactions forbids
fermion masses to appear.
\end{description}
This last point is readily seen. Recall that,
under SU(2)$\times$U(1),~{\rm for
example}
$\pmatrix{\nu_e\cr e\cr}_{\rm L}$ forms a doublet while $e_R$ is
a singlet. Obviously, then, having an explicit electron mass,
\[
{\cal{L}}_{\rm mass} \ = \ m\overline{e}e
\ = \ m\overline{e}_{\rm L}e_{\rm R}
+ m\overline{e}_{\rm R}e_{\rm L},
\]
 would violate SU(2) symmetry.

Although the asymmetric $SU(2)$ assignments for electrons prevent the
above mass term,
it is not quite true that
all mass terms are forbidden by the nature of the weak interactions.
One must really distinguish between
two types of mass terms:  Dirac masses,
and Majorana masses.  A Dirac mass is of the form
\[
{\cal L}_{\rm mass}^{\rm Dirac}
= m_D(\overline{\psi}_{\rm L}\psi_{\rm R}
+ \overline{\psi}_{\rm R}\psi_{\rm L})\, .
%\hspace*{18em}
\]
Clearly $m_D$ is forbidden by the $SU(2)$ assignments of fermions in
the {\sc sm}, since
$\psi_{\rm L}$ is a doublet, while $\psi_{\rm R}$ is a singlet.
In addition, however, one
can have Majorana masses which involve
\[
{\cal L}_{\rm mass}^{\rm Majorana}
= \frac{m_{\rm R}}{2}\left(\psi_{\rm R}^{\rm T}\,C\psi_{\rm R}
+ \overline{\psi}_{\rm R}\,C\overline{\psi}_{\rm R}^{\rm T}\right)
+ \frac{m_{\rm L}}{2}\left(\psi_{\rm L}^{\rm T}\,C\psi_{\rm L}
+ \overline{\psi}_{\rm L}\,C\overline{\psi}_{\rm L}^{\rm T}\right),
\]
\noindent
where $C$ is the charge conjugation matrix.  In the {\sc sm},
$m_{\rm L}$ is forbidden by
$SU(2)\times U(1)$, for all fermions.  However,
$m_{\rm R}$
is allowed by $SU(2)$ and by $U(1)$ for the right-handed
neutrinos, $\nu_R$,  since right-handed
neutrinos have no $SU(2)\times U(1)$ quantum numbers at all.
Therefore it is possible to have a Majorana mass term in the {\sc sm} of
the form
\[
{\cal L}_{\rm mass}^{\rm Majorana} = \frac{m_{\rm R}^\nu}{2}
\left\{\nu_{\rm R}^{\rm T}\,C\nu_{\rm R}
 + \overline{\nu}_{\rm R}\,C\overline{\nu}_{\rm R}^{\rm T}
\right\}.
\]
Thus, more correctly, $m^{\nu}_{\rm R}$ is the only possible explicit
mass parameter in the {\sc sm}. \footnote{To be more precise,
\protect $m^{\nu}_{\rm R}$ is a \protect$3 \times 3$ mass matrix,  not a
single mass parameter.}  It could very well
be zero but, as we shall argue later, is probably very large.  At any
rate, this parameter does not appear to play a fundamental role
except for neutrinos.

Even though ${\cal L}_{\rm SM}$ has no explicit mass scales in it,
besides $m^{\nu}_R$,  two
dynamical scales enter in the theory whose origins are
completely different. These scales are
$\Lambda_{\rm QCD}$ - the dynamical scale associated with
the unbroken color $SU(3)$ theory - and $v_F$, the Fermi scale -
generated as the result of the spontaneous breakdown of the $SU(2)\times
U(1)$ electroweak group down to $U(1)_{em}$.  Let me briefly discuss, in
turn, the origins of these two fundamental scales.

\subsection{The origin of $\Lambda_{\rm QCD}$}

Although coupling constants are dimensionless, they are not real
constants but depend on the momentum scale where they are measured.
Examining corrections at higher order, one sees that the effective
coupling constant in {\sc qcd} is a running coupling constant.  This
behaviour arises from quantum corrections to Green's functions in field
theory. Green's functions contain, in general, infinities when
computed perturbatively. These infinities can be removed
by rescaling the fields and couplings,
\[
A^\mu = \sqrt{Z_3}\,A^\mu_{\rm Ren} \, ,
\qquad g_3 = Z_g (g_{3,\rm Ren})  \, ,
\]
and specifying the value of certain Green's functions at some
normalization point.  For example, in QCD one fixes the three-gauge
vertex of Fig. 2 to have the strength $Z_g^{-1}Z_3^{-3/2}$
at $p^2 = q^2 = r^2 = \mu^2$ - a convenient normalization point.
\begin{figure}
\vspace{6cm}
\caption[]{Three-gauge vertex in~{\rm QCD}.}
\end{figure}
\noindent
Physical amplitudes should not depend on the arbitrary normalization
point $\mu^2$ which one chooses.  Although $\mu^2$ is arbitrary, the
response of the theory to $\mu^2$ changes gives physical information.
This information is codified in the Renormalization Group Equations
\cite{RGE}.

The {\sc rge} arise from studying the $\mu^2$-dependence of the Green's
functions in the theory. The relationship between
the renormalized (finite) Green's functions
and the original Green's functions is $\mu^2$-dependent.
For instance, the renormalized gluon propagator in {\sc
QCD} is related to the original propagator by the equation
\[
 G_{\rm Ren}^{(2)}(\mu^2)= Z_3^{-1}(\mu^2)\,G^{(2)} \ .
\]
The dependence of $G_{\rm Ren}^{(2)}$ on $\mu^2$ enters both explicitly
and
through the coupling
constant dependence on $\mu^2, g_{3, {\rm Ren}}(\mu^2)$.
Since $G^{(2)}$ is independent of $\mu$, it follows that
\[
\left[\mu\frac{\partial}{\partial\mu}
+ \beta(g_{\rm 3, Ren})\frac{\partial}{\partial g_{\rm 3, Ren}}
- \mu\frac{\partial}{\partial\mu}Z_3^{-1}(\mu)\right]
G_{\rm Ren}^{(2)}(\mu^2) \ = \ 0 \ .
\]
This is the {\sc RGE} for the gluon propagator.
Here $\beta(g_{\rm 3, Ren})$ describes how the coupling
depends on the normalization point $\mu^2$:
\[
\beta(g_{\rm 3, Ren}) = \mu\frac{\partial}{\partial\mu}
g_{\rm 3, Ren}(\mu^2)\ .
\]
Thus, the $\beta$-function effectively
fixes the way in which the Green's functions depend on $\mu^2$.
$\beta$ can be calculated perturbatively from the defining equation,
using that
\[
g_{\rm 3, Ren}(\mu^2) =\frac{1}{Z_g(\mu^2)}\,g_3
\]
along with a perturbative expansion of $Z_g$ in $g_3$.

Because {\sc qcd} has no intrinsic scale, the $\mu^2$ dependence is
related to the $q^2$ dependence of the propagator:
\[
G_{\rm Ren}^{(2)}[\mu^2] \equiv G_{\rm Ren}^{(2)}
[q^2/\mu^2; g_{3,~{\rm Ren}}(\mu^2)] \ .
\]
In turn,  the momentum dependence of the Green's functions is related
to the scale dependence of the effective coupling
$\overline{g}_3(q^2)$.  This effective coupling constant - or running
coupling constant - is defined through the equation \cite{RGE}
\[
\left[q^2\frac{\partial}{\partial q^2}
- \beta(g_{\rm 3, Ren})\frac{\partial}{\partial g_{\rm 3, Ren}}
\right]\overline{g}_3(q^2) \ = \ 0 \ ,
\]
along with the boundary condition that
the effective coupling specified at some scale, $q_0^2$, is
$g_{3,~{\rm Ren}}$.  That is,
\[
{\overline{g}}_3 (q_0^2) =
g_{3,~Ren} \ .
\]
\noindent
In {\sc QCD} the $\beta$-function $\beta (g_{3, Ren})$
can be computed perturbatively and one finds \cite{GW}
\[
\beta(g_{\rm 3,~ Ren}) = -b_3\,\frac{g^2_{3,\ Ren}}{16\pi^2}
+ \cdots,
\]
where
\[b_3 = \left[11 -\sfrac{2}{3}N_{\rm quarks}\right].
\]
Using this result, the running coupling constant squared for QCD
is seen to be
\[
\alpha_3(q^2) \equiv \frac{{\overline{g}}_3^2(q^2)}{4\pi}
= \frac{\alpha_3(q_0^2)}
{\dst 1 + \frac{b_3}{4\pi}\alpha_3(q_0^2)
\ln\left(\frac{q^2}{q_0^2}\right)}.
\]

The running coupling constant squared $\alpha_3 (q^2)$ given by the
above equation is \\ sketched in Fig. 3. The running of this coupling
constant with $q^2$ describes qualitatively two important features of
the strong interactions:
\begin{figure*}
\vspace{6cm}
\caption[]{Behaviour of the {\rm QCD} running coupling constant with
$q^2$.}
\end{figure*}

\begin{description}
\item{i)} We do not observe free quarks and gluons but only hadrons
({\bf confinement}), presumably because at small values of
$q^2$ the coupling constant
becomes strong.

\item{ii)} At short distances---large $q^2$---hadrons appear to be
composed of essentially free quarks ({\bf asymptotic freedom}).
\end{description}

We can define $\Lambda_{\rm QCD}$ as the scale where
$\alpha_3 \approx 1$. Clearly, this is
a truly dynamical scale.
A measurement of this scale, from deviations at high $q^2$
of deep inelastic scattering of hadrons from free quark behaviour,
gives \cite{DIS}
\[
\Lambda_{\rm QCD} \sim \mbox{200--300\,MeV}.
\]
In principle, when the confinement problem is finally calculationally
under control, we should be able to compute hadron masses (protons,
pions, \etc) directly in terms of
$\Lambda_{\rm QCD}$ \footnote{This will not be
quite true for heavy quarks, since there the
masses of the heavy quarks contribute non-negligibly to the
corresponding hadronic masses.}.

\subsection*{The origin of the Fermi Scale $v_F$}

Although SU(2)$\times$U(1) is a good symmetry of ${\cal L}_{\rm SM}$,
this symmetry is not respected by the vacuum, leading to the
symmetry breakdown:
\[
\mbox{SU(2)}\times\mbox{U(1)} \to \mbox{U(1)}_{\rm em}.
\]
 This breakdown is associated with a physical scale,
the Fermi scale $v_{\rm F}$.
All the masses of elementary excitations in the standard model (quarks,
leptons, and gauge bosons) are proportional to $v_{\rm F}$, with
\[
v_{\rm F} = \left(\sqrt{2}\,G_{\rm F}\right)^{-1/2}
\approx 250\mbox{GeV}.
\]
 The role of symmetry breakdown as a generator of mass scales is
familiar in superconductivity where a Cooper pairing between opposite
spin electrons, $\big<\psi\uparrow\,\psi\downarrow\big>$,
leads to a mass gap between the
normal and the superconducting
ground states\cite{BCS}.  Similarly, here
$v_{\rm F}$ is the scale of the order
parameter of the breakdown of $SU(2)\times U(1) \to U(1)_{\rm
em}$.  One of
the most crucial questions in the electroweak theory is ``what is
the origin of this order parameter?"

For the electroweak theory two alternatives have been proposed for the
Fermi scale.  The simplest of these is that:

\begin{description}
\item{i)} The Fermi Scale $v_{\rm F}$ is connected to the vacuum expectation
value (\rm VEV) of an elementary scalar field.
\end{description}
\noindent
Alternatively, it has been proposed that:
\begin{description}
\item{ii)} The Fermi Scale $v_F$ is associated with the scale of some
dynamical condensate of fermions from some deeper theory.
\end{description}

\noindent As we shall see in what follows, the physical consequences of
$v_{\rm F}$
being the {\rm VEV} of some elementary scalar field are quite different than
those of having $v_{\rm F}$ being associated with some underlying fermion -
antifermion condensate, $\langle\bar{F} F\rangle$.

For SU(2)$\times$U(1) there is, however, another perhaps deeper mystery.
Although both the gauge boson masses and the masses of all fermions are
proportional to $v_{\rm F}$:
\[
M_W \sim v_{\rm F};\qquad\qquad m_{\rm f} \sim v_{\rm F} \ \ ,
\]
there is no real understanding why $M_W \sim 100\,$GeV, while most fermion
masses (except for the top!) have $m_{\rm f} \ll v_{\rm F}$.
Furthermore, there is also  no deep understanding of the hierarchies
among fermion masses.  That is why, for example, $m_\tau$ : $ m_\mu$ : $m_e$
is approximately in the ratio of 3500 : 200 : 1.

In the simplest version of the {\sc sm}\cite{Weinberg} the possibility
of inducing the breakdown
\[
\mbox{SU(2)}\times\mbox{U(1)} \to \mbox{U(1)$_{\rm em}$}
\]
is effected through the introduction of an
elementary self coupled scalar field,
$\Phi$, which is a complex $SU(2)$ doublet.  This field is assumed to
have an asymmetric potential

\[ V \ = \ \lambda\left(\Phi^{\dagger}\Phi
-\frac{v_{\rm F}^2}{2}\right)^2\ ,
\]
\noindent
which leads to the symmetry breakdown. The gauge boson masses,
$M_W$ and $M_Z$, arise from the
couplings of this scalar field to the $SU(2) \times U(1)$ gauge
fields.  The replacement of $\Phi$ by its ${\rm VEV}, < \Phi > = v_{\rm
F}/\sqrt{2}$, in the seagull graph of Fig. 4 gives
\[ M_{\rm w} = \frac{g_2 v_{\rm F}}{2} \, .
\]
\noindent
Since the $SU(2)$ coupling constant $g_2$ is `known', $M_W$ is fixed
directly from a knowledge of the Fermi scale,  $v_{\rm F}$.
\begin{figure}
\vspace{6cm}
\caption[]{Seagull graph which gives the $W$ bosons a mass when $\Phi$
is replaced by its VEV.}
\end{figure}

The fermion masses also originate from $\Phi$,
but they require new couplings. These are the, so-called,
Yukawa couplings $\Gamma_f$.  These couplings connect  by means
of the doublet Higgs field $\Phi$,
the $SU(2)$
singlet right-handed fermions with the left-handed fermions which are in
doublets.   When $\Phi$ gets
replaced by its {\rm VEV} this leads to a mass term.
\[
 {\cal{L}}^{\rm fermion}_{\rm mass} =
  \dst \frac{(\Gamma_f\,v_{\rm F})}{\sqrt{2}}
(\overline{f}_{\rm L}f_{\rm R}
+ \overline{f}_{\rm R}f_{\rm L})
\]
\noindent
Even though all fermion masses are proportional to $v_{\rm F}$, they
are not determined in the standard model
because the Yukawa couplings $\Gamma_f$ are `unknown'.
Since the $\Gamma_f$ are arbitrary, any prediction for
the fermion masses clearly requires physics beyond the {\sc sm}.

I will return to the question of fermion masses
later on, but first I want to ask
another question:
`what physics fixes the scale $v_{\rm F}$?'
By itself, since $v_{\rm F}$ is the only scale in
the standard model, the question makes no sense.
However, if we consider the {\sc sm} within a broader context,
as an effective theory up to some cut-off $\Lambda$
(\eg $\Lambda = M_{\rm Planck}$), then it makes sense
to ask for the relationship between $v_{\rm F}$ and
$\Lambda$. Indeed, because the $\lambda \phi^4$ theory is
`trivial'\cite{trivial}\footnote{
\ie the only consistent version
of the theory has $\lambda_{\rm Ren} = 0$.},
considering scalar interactions in the {\sc sm} without some
high energy cut-off is itself questionable.

In a $\lambda \phi^4$ theory - unlike what happens in {\rm QCD} - the
effective coupling constant $\lambda (q^2)$ increases with increasing
$q^2$.  The solution of the {\rm RGE}
\[ \frac{d \lambda}{d\ell n q^2} \quad= \frac{3 \lambda^2}{4
\pi^2} + \cdots \ \ ,
\]
neglecting the higher order contributions, is
\[
\lambda(q^2) = \frac{\lambda(\Lambda_0^2)}
{\dst 1 - \frac{3\lambda(\Lambda_0^2)}{4\pi}
\ln\left(\frac{q^2}{\Lambda_0^2}\right)}
\]
which blows up at the
location of the Landau pole \cite{Landau}.
\[
\Lambda_c^2 = \Lambda_0^2\,e^{\dst 4\pi^2/3\lambda(\Lambda_0^2)} \, .
\]
The behaviour of $\lambda(q^2)$ versus $q^2$ is sketched in fig. 5. As
one can see from this figure,
\begin{figure}
\vspace{6cm}
\caption[]{Behaviour of the scalar effective coupling constant as a
function of $q^2$.}
\end{figure}
once the location of the Landau pole is known (in practice, once one
knows where the
coupling constant $\lambda$ becomes large) then one knows
the value of $\lambda(q^2)$ for all $q^2$. Furthermore, it is clear that
the
theory is perfectly sensible for
$q^2 \ll \Lambda_c^2$. Note that if one insists that the Landau pole
is pushed to infinity,  then $\lambda (\Lambda_0^2) \to 0$.  This is
a qualitative way to see the import of the
statement of triviality.

In the electroweak theory, since the scalar doublet $\Phi$ is self
interacting, one expects to
have a very similar situation to the one just described \cite{LW}.
Physically,
the scalar sector makes perfect sense, as long as we assume that there
is some physical high energy cut-off.
We can `measure' where the cut-off is in the scalar theory
from the value of the Higgs mass - the remaining physical scalar
excitation in the theory with just one scalar doublet $\Phi$.
Using the potential $V$, one finds for this mass the formula
\[
M_{\rm H}^2 = 2\,\lambda(M_{\rm H}^2)\,v_{\rm F}^2.
\]
{}From this formula one sees that as long
as the Higgs is light, the cut-off is very far away.
In fact $\Lambda_c \approx 10^{19}\,$GeV
$\equiv M_{\rm Planck}$ would be obtained for
$M_{\rm H} \approx 200\,$GeV. \footnote{For reference,
the {\sc lep} limit on the Higgs mass is $M_{\rm H} > 60\,$GeV.\cite{bound}.}
 Thus, in this case, the scalar effective field theory is very good
and it is sensible to ask whether having
$v_{\rm F} \ll \Lambda_c$
is a stable condition ({\bf naturalness condition}.)
If, on the other hand, the Higgs is very heavy,
so that $M_{\rm H}$ and $\Lambda_c$ are comparable,
then the effective field theory for the scalar sector
stops making sense.
Numerically, this occurs around \cite{LW}
\[
M_{\rm H} \sim \Lambda_c \sim \mbox{800--1000\,GeV}.
\]
 In this situation, $\left<\Phi\right>$
as an SU(2)$\times$U(1) order parameter must be
replaced by something else.

{}From the above discussion, one sees that if
the Higgs is very light one must worry about the
naturalness of the hierarchy
\[
v_{\rm F} \ll \Lambda_c .
\]
In fact, radiative effects in a theory with a cut-off
destabilize this hierarchy.
For instance, the Higgs mass gets a quadratic mass
shift and so there is a similar shift, in $v_{\rm F}$

\[ v_{\rm F}^2 \to
v_{\rm F}^2 + \left[\frac{\alpha}{\lambda}\right]\Lambda_c^2
\]
The existence of quadratic radiative shifts for scalar fields is
problematic, for to keep
the hierarchy
$v_{\rm F} \ll \Lambda_c$
requires an enormous amount of fine tuning.  This fine tuning
can only be avoided if:

\begin{description}
\item{i.)} Instead of $v_F \ll \Lambda_c$, $\Lambda_c \sim v_{\rm F}$.
In this case, however, one abandons the notion
of the Fermi scale coming from an
elementary scalar field.
\item{ii.)} There is some protective symmetry which prevents large
radiative shifts.
\end{description}

Remarkably, such a protective symmetry exists; it is called
supersymmetry \\ ({\sc susy})\cite{susy}.  This is a boson-fermion
symmetry in which bosonic degrees of freedom are
paired up with fermionic degrees of freedom, with
$m_{\rm Boson} = m_{\rm Fermion}$
in the limit of exact {\sc susy}. In a
{\sc susy} version of the standard model the
quadratic divergences cancel.
Even with broken {\sc susy} the divergences are ameliorated.
For instance, the quadratic Higgs mass shift due to the presence of
a $W$-loop in a {\sc SUSY} version of the {\sc SM} gets cancelled by the
equal and opposite shift due to the presence of a Wino-loop $(\widetilde W)$.
Here $\widetilde{W}$ is the spin-1/2 partner of the $W$.
These two graphs are shown in Fig. 6.
\begin{figure}
\vspace{6cm}
\caption[]{Graphs contributing to the Higgs mass shift in a {\rm SUSY}
version of the {\sc sm}.  The Wino graph, since it involves fermions,
has an overall sign of -1 relative to the $W$-loop graph.}
\end{figure}
Obviously, {\sc susy} is not an exact symmetry of nature,
since no superpartners have yet been seen for any of the
known particles.
However {\sc susy} has no need to be exact to ensure naturalness.
Indeed even if ${\widetilde{M}}_{\tilde W} \not= M_{W}$, the presence of
the Wino-loop now changes the cut-off dependence of the shift to be
only logarithmic:
\[
v_{\rm F}^2 \to v_{\rm F}^2 + \frac{\alpha}{\lambda}
({\widetilde{M}}^2_{\tilde W} - M_W^2)
\ln\left(\frac{\Lambda_c}{v_{\rm F}}\right).
\]
To guarantee naturalness, therefore, it suffices
that the {\sc susy} mass splitting be not
too large. That is,
\[
(\widetilde{M} - M) \leq  O(v_{\rm F}).
\]

Thinking about $v_{\rm F}$ has brought us to consider
two interesting alternatives for the nature of the order parameters for
the breaking of $SU(2)\times U(1) \longrightarrow U(1)_{\rm em}$.  These
alternatives lead to quite different physics.  In particular, if
\begin{description}
\item{i.)} the breaking is through an elementary scalar {\sc vev}
$\left<\Phi\right>$,
then naturalness suggests that one should have both a light
Higgs and supersymmetry.
\end{description}
\noindent
On the other hand, if
\begin{description}
\item{ii.)} the breaking is dynamical, then $v_{\rm F}$
is a scale like $\Lambda_{\rm QCD}$ and one expects to have a new strong
interacting theory in the neighborhood of $1~${\rm TeV}.
\end{description}
In what follows, I will discuss these two alternatives in some more
detail.

%%%%%%%%%%%%%%%%%%%%%%%%%
%%%%%%%%%%%%%%% SECTION 2
%%%%%%%%%%%%%%%%%%%%%%%%%
%\input{parttwo.tex}
%> $.PECCEI.parttwo
\section{The SUSY Alternative}

It is relatively straightforward to write down a supersymmetric
version of the {\sc sm}\cite{MSSM}.  For these purposes one must:
\begin{description}
\item{i.)} Associate scalar partners to quarks and leptons, building a
chiral supermultiplet
\[
\Phi = \{\phi\,;\,\psi\} \, .
\]
In the above, the scalar supermultiplet $\Phi$ is composed of a
complex scalar $\phi$ and a Weyl fermion $\psi$.  For example,
the left-handed electron supermultiplet is given by

\[
\Phi_{e_L} = \{\widetilde{e}_{\rm L}\,;\,e_{\rm L}\} \, ,
\]
where $\widetilde{e}_{\rm L}$ is a selectron
and $e_{\rm L}$ is the left-handed electron.
Similarly, the Higgs doublet (two are actually needed---
see below) is the scalar component of a chiral
supermultiplet.
\item{ii.)} Associate spin-1/2 partners to the gauge fields,
building vector supermultiplets
\[
V = \{ V^\mu\,;\,\lambda\} \, .
\]
Here $V^\mu$ is the gauge field and $\lambda$ the
spin-1/2 gaugino, which is a Weyl fermion.

\item{iii.)} Supersymmetrize all interactions; \eg for Yukawa
coupling one has now the three graphs of Fig. 7.
\end{description}
\vbox{
\vspace{7cm}
%\begin{figure*}[h]
%\vspace{6cm}
\begin{center}
{\bf Figure 7.} {\it Yukawa vertices
entering in a {\sc susy} version of the {\sc sm}.
The fields with a tilde are the supersymmetric partners
of the appropriate {\sc sm} fields.}
\end{center}}
%\end{figure*}

In addition {\sc susy} imposes a few constraints on the allowed interactions:

\begin{enumerate}
\item[i.)] Interactions involving the chiral supermultiplets
should be derivable from a
superpotential $W(\Phi_i)$ which involves only $\Phi_i$
but not $\Phi_i^{\dst\ast}$.
This circumstance requires the presence of two
different Higgs supermultiplets in the {\sc susy sm}
since the vertex diagram of Fig. 8 is not allowed.
To generate the $u$-quark Yukawa couplings one
replaces $H_1^*$ by a new Higgs doublet $H_2$, which is
itself part of a chiral supermultiplet.
\footnote{The need for introducing a second Higgs supermultiplet
in the {\sc susy sm}
can also be motivated by demands
that there be no U(1) gauge anomaly.
The absence of this anomaly requires
${\rm Tr}[Q]_{\rm fermions}=0$.
Introducing $\widetilde{H}_1$ with $Q_{\widetilde{H}_1}=(0,+1)$
immediately requires a second fermion Higgs, $\widetilde{H}_2$,
with $Q_{\widetilde{H}_2}=(0,-1)$ in order to ensure anomaly
cancellation.}

\item[ii.)] The scalar potential in the theory follows from
$W(\Phi_i)$ plus other terms (the, so called, D-terms)
arising from supersymmetrizing the gauge interactions.
One finds:
\[
V(\Phi) = \sum_i
\left|\frac{\partial W}{\partial\Phi_i}\right|^2
+ \frac{g_a^2}{2}\left|\Phi_i^{\dst\ast}T_a\Phi_i\right|^2
\]
where $T_a$ is a group generator.
\end{enumerate}

\vbox{
\vspace{7cm}
%\begin{figure*}[h]
%\vspace{6cm}
%\caption[]{Vertex diagram forbidden by {\sc susy}.}
%\end{figure*}
\begin{center}
{{\bf Figure 8.} \it Vertex diagram forbidden by {\sc susy}.}
\end{center}}

These points are of importance for the issue of
$SU(2)\times U(1)$ breaking.
Ignoring Yukawa couplings, the superpotential only
contains a term
\[
W(H_1,H_2) = \mu\,H_1H_2 \\ ,
\]
where $\mu$ is a scale parameter,
because $(H_1H_2)^2$ terms in the superpotential
would result in a non-renormalizable
$V(\Phi)$.
Thus, including D-terms but neglecting Yukawa interactions,
one has for the potential
\begin{eqnarray*}
V(H_1, H_2) & = & \frac{1}{8}g_2^2
\left(H_1^{\dst\ast}\vec{\tau}H_1 + H_2^{\dst\ast}\vec{\tau}H_2\right)
\cdot\left(H_1^{\dst\ast}\vec{\tau}H_1 + H_2^{\dst\ast}\vec{\tau}H_2\right)\\
&& + \frac{1}{2}g_1^2
\left(H_1^{\dst\ast}H_1 - H_2^{\dst\ast}H_2\right)^2
+\mu^2\left(H_1^{\dst\ast}H_1 + H_2^{\dst\ast}H_2\right)
\end{eqnarray*}
Note that in the {\sc susy sm} the quartic coupling constants $\lambda$ are
related to the gauge couplings.
Since $\mu^2 > 0$, the above potential
cannot break SU(2)$\times$U(1) if {\sc susy} is exact.
However, one must still add {\sc susy} breaking terms and there is
the possibility that the mechanism which
breaks {\sc susy} might also break SU(2)$\times$U(1).

A popular scenario is to assume that {\sc susy}
is broken at a high scale by a hidden sector,
coupled only gravitationally to the real world.\cite{Sugra}
This scenario introduces only a few parameters
($m_{1/2}$, $m_0$, $A$, $B$)
typifying {\sc susy} breaking and is reasonably predictive.
One can show that this ``hidden sector breaking" leads to a
Lagrangian given by:

%\newpage

\begin{eqnarray*}
{\cal L} & = & {\cal L}_{\rm SM}^{\rm SUSY}
- \sum_i m_0^2\phi_i^{\dst\ast}\phi_i
-\sum_i m_{1/2}\lambda_i\lambda_i
+ A\,m_0W_3(\phi) + B\,m_0W_2(\phi) + \mbox{h.c.}
%&&\hspace*{3em}\mbox{scalar fields}
%\hspace*{2em}\mbox{gauginos}
\end{eqnarray*}
Here $\phi_i$ are all the scalar fields in the theory,
while the $\lambda_i$ are the gauginos.
$W_3(\Phi)$ is the tri-linear superpotential
and $W_2(\Phi)$ is the bi-linear superpotential.

Ignoring Yukawa couplings again, the potential one derives
from this Lagrangian is
\begin{eqnarray*}
V(H_1, H_2) & = & \frac{1}{8}g_2^2
\left|H_1^{\dst\ast}\vec{\tau}H_1 + H_2^{\dst\ast}\vec{\tau}H_2\right|^2
+ \frac{1}{2}g_1^2
\left|H_1^{\dst\ast}H_1 - H_2^{\dst\ast}H_2\right|^2\\[4pt]
&& + \mu_1^2\,H_1^*H_1+\mu_2^2\,H_2^*H_2
-\mu_3^2\,\left(H_1H_2+H_1^*H_2^*\right),
\end{eqnarray*}
\noindent
where $\mu_1^2=\mu_2^2=(\mu^2+m_0^2)$; $\mu_3^2=B \ m_0\mu$.
For SU(2)$\times$U(1)$\to$U(1)$_{\rm e.m.}$ breaking
we need a negative eigenvalue in the Higgs mass matrix
\[
M^2_H = \pmatrix{\mu_1^2&-\mu_3^2 \cr
-\mu_3^2&\mu_2^2\cr}
\]

\vbox{
\vspace{7cm}
%\begin{figure*}[h]
%\vspace{6cm}
\begin{center}
{{\bf Figure 9.} \it Top and stop loops contributing to the
running of $\mu^2_2(q^2)$.}
\end{center}}
%\end{figure*}
\noindent
Now
{\rm det} $M^2_H < 0$
implies that
$\mu_1^2\mu_2^2 < \mu_3^4$.
However, if as here, $\mu_1^2 = \mu_2^2$,
then one has to worry about the stability of $V$, when
$\mu_1^2 < \mu_3^2$.
In fact, there is really no problem at all
since also the masses evolve with $q^2$
and we only expect $\mu_1^2 = \mu_2^2$ at the scale of
{\sc susy} breaking.
Indeed one can argue that, since the top quark is so heavy,
$\mu_2^2$ changes sign as one evolves down from high to low scales.\cite{IR}
In effect, {\sc susy} breaking triggers the breaking of
SU(2)$\times$U(1)$\to$U(1)$_{\rm e.m.}$!

%The following are examples, from Rocky Kolb's lectures

Let us try to understand this phenomena a bit more in detail.
The reason why $\mu_2^2$ evolves much more rapidly than
$\mu_1^2$ with $q^2$ is
solely due to the presence of a heavy top quark.  Roughly speaking, for $q^2$
near the {\sc susy} breaking scale $M^2_{SB}$,
the loops containing top and stop
shown in Fig. 9 are dominated predominantly by the stop contribution, since
$m_t\simeq 0$ at these large scales while $\tilde m_{\tilde t}\simeq
m_0\simeq \mu_2$.
One finds\cite{Kyushu}
\[
\delta\mu_2^2=-\frac{3}{4\pi^2}h_t^2\mu_2^2
\ln[M_{\rm SB}^2/q^2] \, .
\]
\noindent
These contributions dominate over that of all the other quarks and
leptons, because the top Yukawa coupling, $h_t$, is so much bigger than the
rest.  Furthermore, the evolution of $\mu_2^2$ is rapid enough so that indeed
for $q^2\sim M^2_W\, , \,\, \mu_2^2(q^2)$ becomes negative triggering the
electroweak breakdown.
A schematic behavior of what happens is shown in Fig. 10.

\vbox{
\vspace{7cm}
%\begin{figure*}[h]
%\vspace{6cm}
\begin{center}
{{\bf Figure 10.} \it Different evolution of $\mu_1^2$ and $\mu^2_2$
with $q^2$.}
\end{center}}
%\end{figure*}

Although the above is a very interesting scenario for generating the Fermi
scale, it also illustrates that while
{\sc susy} solves the naturalness problem, it
leaves unexplained a hierarchy problem.  Because $\mu_2^2$ evolves only
logarithmically with $q^2$, the Higgs mass is directly related to the {\sc
susy} breaking mass parameter $m_0^2$
\[
M_H^2\simeq -\mu_2^2(M^2_W) \sim \mu^2_2(M^2_{{\rm SB}})
\simeq m_0^2
\]
\noindent That is, in magnitude there cannot be much difference between the
value of $\mu_2^2$ at high and low scales.  The deep unanswered question is
how then at the very large scale where {\sc susy} breaking presumably takes
place, $M_{{\rm SB}}$, can one generate such a small mass as $m_0$--which
is in
the 100 GeV-1 TeV range.  What is the reason for the hierarchy
\[
M_{SB} \gg m_0 \, ?
\]

I should remark that, in effect, the radiative breaking due to stop and
top loops evolves even more rapidly than one may think, since the Yukawa
coupling $h_t$ is also $q^2$ dependent and--like the Higgs coupling $\lambda$--
it is not asymptotically free.  In fact, the actual structure of the RGE for
$h_t$ is quite interesting in its own right,
since the inclusion of strong interaction effects
forces $h_t$ to a fixed point.\cite{PR}  One has
\[
\frac{dh_t}{d\ln q^2}=\frac{h_t}{16\pi^2}\biggl\{6h_t^2-\frac{16}{3}
g_3^2\biggr\} \, .
\]
\noindent Even though for large $h_t$ the effects of QCD are negligible, the
above equation forces $h_t$ to evolve to a value given by the vanishing of the
curly bracket.  For $q^2\simeq m_t^2$, this fixed point value
\[
h_t^*\simeq \frac{4}{3} \sqrt{2\pi\alpha_s(m_t^2)}
\]
is numerically very close to unity and leads to an upper bound
for the top mass.

In the
minimal {\sc susy} standard model the
top quark mass, $m_t$, is related
to $\bigl<H_2\bigr>=v_2/\sqrt{2}$, while the Fermi scale, $v_F$, measures
both $\bigl< H_2\bigr>$ and $\bigl<H_1\bigr>=v_1/\sqrt{2}$.  From the
Higgs potential $V(H_1,H_2)$, one finds
\[
v_F^2=v_1^2+v_2^2 \equiv v_F^2 \cos^2\beta+v_F^2\sin^2\beta
\]
\noindent leading to
\footnote{A more precise {\sc RG}
evolution gives $m_t(m_t^2)\simeq 190\sin\beta \,$
GeV.\cite{BBO}}

\[
m_t=\frac{h_t^*}{\sqrt{2}}v_2=h_t^* \frac{v_F}{\sqrt{2}}
\sin\beta \simeq 175 \sin\beta \, GeV \, .
\]
\noindent
This value is compatible for reasonable
$\sin\beta$, with the direct limit on the top obtained at
Fermilab.\cite{Barbieri}
A recent analysis of the {\sc susy sm} radiative
corrections\cite{Babrieri} yields a best value for the top mass of
\[
m_t=131^{+31}_{-23} \, GeV \, ,
\]
\noindent which, taken in conjunction with the above {\sc RGE} result, gives
\footnote{Note that, for all practical purposes the result of radiative
corrections in the {\sc susy sm} is equivalent to that of the SM with a light
Higgs.}
\[
\sin\beta=0.68 \pm 0.16 \qquad (0.6 < \tan\beta < 1.5)
\]

There are three remarkable results which follow from a {\sc SUSY} extension of
the {\sc sm}:

(i) {\sc susy} requires a larger Higgs sector and a quite light Higgs boson is
expected.  The two Higgs doublets necessitated by a {\sc susy} extension $H_1$
and $H_2$ contain 8 real fields, 3 of which are absorbed to give masses to the
$W^+$, $W^-$ and $Z^0$.  Thus we are left with 5 physical fields --- 2 charged
$(H^+ \,,H^-)$ and 3 neutral $(h^0,\,H^0,\,A)$ Higgses.  The potential
$V(H_1,\,H_2)$ leads to a number of tree-level relations among these states and
the masses of the gauge bosons.  One finds:\cite{HHG}
\begin{eqnarray*}
M_H^2 & = & M_W^2 + M_A^2\\
M_h^2 + M_H^2 & = & M_Z^2 + M_A^2\\
M_h & \leq & M_A \ \leq \ M_H\\
M_h & \leq & M_Z|\cos2\beta|
\end{eqnarray*}

For the value of $\beta$ given above, the last relation is problematic, since
the lightest neutral Higgs, $h$, has not been observed at LEP.\cite{bound}
However, it turns out that radiative corrections to these equations
are large (due to the
large mass of the top) and can lead to $M_h > M_Z$.  One can appreciate the
order of magnitude of the effect by focusing again on the radiative mass shift
coming from the top and stop loops of Fig. 9.  The shift is proportional both
to the top-stop squared mass difference and to $h_t^2$.  Since the Yukawa
coupling $h_t$ itself is proportional to $m_t$, one sees that the shift due to
radiative effects for the lightest Higgs mass actually depends quartically on
$m_t$.  An approximate formula for the shift\cite{ERZ} is
\[
\delta M_h^2 \simeq \frac{3g_2^2m_t^4}{8\pi^2M_W^2\sin^2\beta}
\ln\biggl[1+\frac{m\tilde q^2}{m_t^2}\biggr]  \, ,
\]
where $m_{\tilde q}$ is a generic squark mass.  For $\cos 2\beta=1$ and
$m_{\tilde q}=1 \, TeV$, this shift is plotted in Fig. 11.  As can be seen, for
large $m_t$, this shift can be of the order of tens of GeVs.

\vbox{
\vspace{9cm}
%\begin{figure*}[h]
%\vspace{8cm}
\begin{center}
{{\bf Figure 11.} \it Lightest Higgs mass
shift as a function of $m_t$.  Adapted from\cite{HH}}.
\end{center}}
%\end{figure*}

A further interesting observation can be made.\cite{Barger}  If one takes the
range of $\beta$ indicated from the value of $m_t$ obtained by radiative
corrections to the {\sc susy sm}, along with the bounds on $M_h$ which follow
from the radiatively corrected version
of the $M_h \leq M_Z|\cos 2\beta|$ relation, as
well as from the direct search bounds from LEP, one obtains an exclusion plot
in the $M_h-m_t$ plane.  This plot is shown in Fig. 12.  As can be seen, it is
conceivable that sharp tests of this scenario will emerge soon
when bounds on
$m_t$ and $M_h$ are sharpened.  At LEPII one should be able to probe value for
$M_h$ up to about 80-90 GeV, while the forthcoming run of the Tevatron should
either discover top or push the top bound to near 160 GeV.

\vbox{
\vspace{9cm}
%\begin{figure*}[h]
%\vspace{8cm}
\begin{center}
{{\bf Figure 12.} \it Exclusion plot in the $M_h-m_t$ plane.
Adapted from\cite{Barger}}.
\end{center}}
%\end{figure*}

(ii) {\sc susy} extensions of the {\sc sm} very naturally provide candidates
for the dark matter in the universe.  This point is easily understood.  Most
simple {\sc susy} extension of the {\sc sm} --- like the trivial one we have
considered above --- preserve a discrete symmetry, known as
$R$-parity.\cite{R}
Roughly speaking, one has $R=+1$ for particles and $R=-1$ for sparticles.
Thus,
as shown in Fig. 7, when one supersymmetrizes an ordinary vertex, one
introduces always {\bf pairs} of sparticles.  $R$-parity of the {\sc susy}
interactions ensures that the lightest supersymmetric particle ({\sc lsp}) is
stable.  Particularly interesting as dark matter candidates is any one of the 4
`neutralinos' --- the {\sc susy} partners of the $(\gamma,\,Z,\,H_1^0,\,H_2^0)$
bosons --- $(\tilde\gamma,\,\tilde Z,\,\tilde H_1,\,\tilde H_2)$.  In general
the {\sc lsp} will be a combination of these states, found by diagonalizing the
neutralino mass matrix and looking for the minimum eigenvalue.  It is beyond
the purpose of these lectures to discuss this problem in detail.  However, I
note that the neutralino mass matrix besides depending on the gauge boson
masses and $\beta$, also depends on the {\sc susy} breaking gaugino mass
$m_{1/2}$ and on $\mu$.  Writing the {\sc lsp} as
\[
\tilde\delta=a_\gamma \tilde\gamma+a_Z\tilde Z+a_1\tilde H_1
+a_2\tilde H_2 \, ,
\]
\noindent one finds that
\[
\tilde\delta=\left\{
\begin{array}{l}
\tilde\gamma \,\, \mbox{as} \,\, m_{1/2}\rightarrow 0\\
\tilde h=\sin\beta \tilde H_1+\cos\beta \tilde H_2 \quad
\mbox{as} \,\,  \mu \rightarrow 0
\end{array}\right.
\]

The reason why neutralinos are excellent candidates for dark matter is because
they have an annihilation rate which is typical of the weak interactions.  The
contribution of weakly interacting massive particles to the universe's relic
abundance
is of order unity, if these particles have an annihilation rate
$\langle\sigma_av\rangle \sim 10^{-37}\, cm^2$\cite{KT}.  The annihilation rate
one computes from the diagram of Fig. 13 is roughly of this order of magnitude,
since the neutralino's fermion-fermion coupling is an electroweak coupling
and, from our naturalness assumption, $m_{\tilde f}$ cannot be that different
from $M_W$ itself.  What is particularly nice is that this diagram not only
determines the LSP abundance, but also sets the rate,
$\tilde\delta+f\rightarrow
\tilde\delta+f$, needed for the detection of neutralinos as dark
matter\cite{Griest}.

\vbox{
\vspace{7cm}
%\begin{figure*}[h]
%\vspace{6cm}
\begin{center}
{{\bf Figure 13.} \it Neutralino annihilation
into fermion-antifermion pairs.}
\end{center}}
%\end{figure*}

(iii) The third interesting property of {\sc susy} is
the way in which the presence of {\sc susy}
matter helps the unification of
the $SU(3)\times SU(2)\times U(1)$ couplings
at high scales.  Using the RGE one can compute the evolution of the coupling
constants in the {\sc sm}.  To one-loop order, one has
\[
\frac{1}{\alpha_i(q^2)}=\frac{1}{\alpha_i(M_Z^2)}+
\frac{b_i}{4\pi} \ln q^2/M_Z^2 \, ,
\]
where the $b_i$ are the coefficients of the $\beta$-function for the particular
coupling in question.  The detailed evolution of the coupling constants depends
on the {\bf matter content} of the theory, since the $b_i$ depend on this
content.  In particular, in {\sc susy} extensions of the {\sc sm} there will be
a different evolution of the $\alpha_i$'s than in the {\sc sm}, since there are
both new fermions (gauginos and shiggs) and new scalars (sleptons and squarks,
as well as now
2 Higgs doublets).

The computation of the contributions of the different particles on the theory
to the $b_i$ coefficients is relatively straightforward.  One
finds\cite{Slansky}
\[
b_i=\sum_{\mbox{\rm particles}}
\biggl[\frac{1}{6} \ell_i^{\mbox{\rm vector}}-
\frac{1}{3} \ell_i^{\mbox{\rm Weyl fermion}}-
\frac{1}{12} \ell_i^{\mbox{\rm real scalar}}\biggr]
\]
\noindent where the $\ell_i$ are group theoretic factors.  For $SU(N)$
groups---which are of interest for our purposes---one has\cite{Slansky}
\[
\ell_i=\left\{\begin{array}{ll}
 2N & \mbox{adjoint representation} \\
 1 & \mbox{fundamental representation}
     \end{array}
 \right.
\]
\noindent For the $U(1)$ group, $\ell_1$ depends on the actual definition of
the charge one takes.  What is used in what follows is
\[
\ell_1 = 2Q_1^2\cdot\frac{3}{5}
\]
\noindent where $Q_1$ is the ordinary $U(1)$ charge in the {\sc sm}
(e.g. $Q_{u_R}=2/3$, $Q_{Q_L}=1/6$; etc.).  As we shall see later on in these
lectures, the extra factor of $3/5$ identifies among all possible $U(1)$
charges the one that sits in a Grand Unified Theory ({\sc gut}).

A straightforward computation leads to the values for the
coefficients $b_i$ for the {\sc sm} and the {\sc susy sm}, shown
in Table 1.
\begin{table}
\caption{Values of $b_i$ in the {\sc sm} and the {\sc susy sm}.}
\begin{center}
\begin{tabular}{|c|c|c|}
\hline Coefficient & {\sc sm} & {\sc susy sm}\\
\hline
$b_1$ & $-41/10$ & $-33/5$ \\
\hline
$b_2$ & $19/6$ & $-1$ \\
\hline
$b_3$ & $7$ & $3$ \\
\hline
\end{tabular}
\end{center}
\end{table}
\noindent
Note that in the {\sc susy sm}, because of the presence of so much
matter, the $SU(2)$ coupling is no longer
asymptotically free.  Using precision
values for the coupling constants obtained
at LEP one can evolve these values
to high $q^2$\cite{Amaldi}.
The result of such a study\cite{Langacker}, using
a {\sc susy} threshold of $10^2-10^3$ GeV, is shown in Fig. 14.
As can be seen, the evolution of the precisely measured couplings at
$M_Z$, exclude the possibility of the values of the couplings meeting at some
large energy scale in the {\sc sm}, but including {\sc susy} modifies the
running so as to rekindle this possibility.  We will return to this point later
on in these lectures.

\vbox{
\vspace{9cm}
%\begin{figure*}[h]
%\vspace{8cm}
\begin{center}
{{\bf Figure 14.} \it Evolution of couplings
without and with {\sc susy} matter.  Adapted from\cite{Langacker}}.
\end{center}}
%\end{figure*}

It may be useful to summarize briefly some of the appealing and interesting
aspects of supersymmetry, as it connects to the question of mass.  We have seen
that introducing {\sc susy} makes the scalar sector of the {\sc sm} natural,
stabilizing the Fermi scale even in the presence of a large cutoff.
Furthermore, {\sc susy} suggests an interesting scenario by which $v_F$ itself
originates as a radiative effect, triggered by the breaking of supersymmetry
through the scalar mass parameter $m_0$.  Why supersymmetry is broken, however,
is not explained.  Furthermore, in effect, a new `naturalness' problem arises
when {\sc susy} breaking is used to try to trigger $SU(2)\times U(1)\rightarrow
U(1)_{em}$.  Namely, why
\[
M_{SB} \gg m_0 \simeq M_W \simeq v_F \, .
\]
\noindent It should be mentioned also that {\sc susy} gives no explanation of
the origin of the magnitudes of the Yukawa couplings for quarks and leptons,
$\Gamma_f$, nor for family replication.  In fact, some further unknown mass
parameters are introduced into the theory with hidden sector breaking: $m_0,\,
m_{1/2}$, as well as the 3 numerical constants $A,\,B$ and $\tan\beta$.  Thus,
in a sense, one has regressed in this area by invoking supersymmetry!
Nevertheless, the overall picture remains quite compelling, particularly if one
can adduce deeper reasons for introducing supersymmetry than naturalness
itself.  As we shall see later on, these reasons exist.

\section{Dynamical Symmetry Breaking}

\noindent The idea behind a dynamical origin for the Fermi scale $v_F$ is
rather simple.  Rather than having the breakdown of $SU(2)\times
U(1)\rightarrow U(1)_{em}$ being caused by an elementary scalar {\sc vev}
$\bigl<\Phi\bigr>$, one replaces
\[
\bigl<\Phi\bigr> \rightarrow
\bigl<\bar FF\bigr> \, ,
\]
\noindent
with $\bigl<\bar FF\bigr>$
being a condensate of fermions from some
underlying theory.
The $\langle\bar FF\rangle$ condensate is assumed to form due to an underlying
gauge interaction, in much the same way as quark condensates
form in {\sc qcd}.
One introduces a new gauge interaction ({\bf Technicolor}\cite{SW})
which is assumed
to become strong at some energy scale $\Lambda_F$.  It follows then, that
\[
\langle\bar FF\rangle \sim \Lambda_F^3
\]
\noindent and that $v_F$ is associated with this dynamical scale.  The picture
is quite akin to {\sc qcd} where the color force becomes strong at scales of
order $\Lambda_{QCD}$ and the quark condensates, which break the global chiral
symmetry of {\sc qcd}, also scale like $\Lambda_{QCD}^3$.

The {\sc sm} value $\rho=1$ which leads to the mass relation
$M_Z^2\cos^2\theta_W=M^2_W$ follows from a custodial $SU(2)$ symmetry of the
Higgs potential (see Aitchison's notes).  One must resort to a similar symmetry
to constrain the technicolor interactions.  One assumes that the technicolor
interactions have an $SU(2)_L\times SU(2)_R \approx O(4)$ invariance which
breaks to $SU(2)_{L+R}\approx O(3)$ as the custodial symmetry leading to
$\rho=1$.  Specifically\cite{SW}, one takes for technicolor, for example, an
$SU(N)$ gauge theory with $SU(2)_L$ doublets
\[
\left(\begin{array}{l}
U_i \\  D_i \\
\end{array} \right)_L
\]
\noindent $i=1,...,N$ and $SU(2)$ singlets $U_{iR}$ and $D_{iR}$.  The
electroweak assignments of $U_i$ and $D_i$ forbid mass terms.  In the absence
of these mass terms and neglecting the weak electroweak forces, the technicolor
theory possess the same kind of chiral symmetry that {\sc qcd} has:  here
$SU(2)_L\times SU(2)_R$.  This symmetry is broken down by condensate formation
\[
\langle\bar U_{Li}U_{Ri}\rangle=
\langle\bar D_{Li}D_{Ri}\rangle \not= 0
\]
\noindent to $SU(2)_{L+R}$, providing the required
custodial symmetry.  The above
condensates also break $SU(2)\times U(1)\rightarrow U(1)_{em}$
and thus serve to give mass to the $W$ and $Z$\cite{SW}.

Although technicolor accomplishes the same things
for the gauge sector as having an
elementary Higgs doublet
---provided $\langle\bar UU\rangle \sim v_F^3$--- the
results one obtains are not quite the same at the level of radiative
corrections.  Furthermore, in simple technicolor, there is no way to generate
Yukawa couplings, $\Gamma_f$, to produce fermion masses.  I will consider both
these points in turn.

Consider first the radiative corrections.  Since the effects of the {\sc tc}
sector are important at scales of order $16\pi^2v_F^2 \gg
M_Z^2$\cite{Chanowitz}, it suffices to
compute the vacuum polarization of the gauge
field propagators in a $q^2$ expansion.
\[
\Pi^{AB}(q^2)=\Pi^{AB}(0) + q^2\Pi^{AB^\prime}(0) + ... \, ,
\]
\noindent with $\{AB\}=\{WW;ZZ;\gamma\gamma;Z\gamma\}$, giving 8 parameters,
$\Pi^{AB}(0)$ and $\Pi^{AB^\prime}(0)$, at order $q^2$.  However,
electromagnetic gauge invariance leads to
\mbox{$\Pi^{\gamma\gamma}(0)=\Pi^{Z\gamma}(0)=0$,} leaving only 6 parameters
to consider.
Three
of these are fixed by the 3 independent parameters of
$SU(2)\times U(1), \,\,g_1,g_2$ and $v_F$ or,
more practically, $\alpha$, $M_Z$
and $G_F$.  Therefore an overall fit of the data---which is independent of
whether or not one uses an elementary Higgs---will be characterized by 3 other
parameters.
These have been called $S,\,T,\,U$\cite{PT} or
$\epsilon_1,\,\epsilon_2,\,\epsilon_3$ in the literature.\cite{ABJ}

In practice,
$S$ (or $\epsilon_3$) is the most interesting parameter, since
$T$ (or $\epsilon_1$) is dominated by information about $m_t$ and
$U$ (or $\epsilon_2$) is not so well determined experimentally, because it
depends sensitively on $M_W$.
If the underlying theory conserves an $SU(2)_{L+R}$---as it
happens in the case of technicolor---then one can write\cite{PT} a useful
spectral representation for $S$ in terms of the functions characterizing the
vacuum polarization tensors of the $SU(2)_{L+R}$ and $SU(2)_{R-L}$ currents in
the theory.  If we denote these vector and axial vector spectral functions by
$v(s)$ and $a(s)$, respectively, one finds\cite{PT}
\[
S=\int^\infty_0 \frac{ds}{s}
[v(s)-a(s)] \, .
\]

Using the above formula,
in simple technicolor models one can estimate $S$ just by scaling up the
spectral functions of {\sc qcd} from $\Lambda_{{\rm QCD}}$ to $\Lambda_F$.
This leads to\cite{PT}
\[
S\simeq a \frac{N}{3} \simeq a \frac{N}{3}
\biggl[\frac{6\pi f_\pi^2}{m_\rho^2}\biggr] \simeq
a \frac{N}{3} 0.25 \, ,
\]
\noindent where, in the second step, one assumes that the spectral function of
{\sc qcd} are dominated by
the $\rho$ (vector) and $A_1$ (axial vector) states.
Here $a$ is the number of technicolor doublets
in the theory and $N$ is the number of technicolors.
Thus in technicolor $S\sim O(1)$ and is positive.  Unfortunately, the combined
analysis of all the electroweak data shown in Fig. 15 gives $S\sim -O(1)$!
However, one need not take this as a catastrophe since the fermion
mass issue, to be discussed below, already suggests that we need a much more
sophisticated technicolor theory than just `scaled up {\sc qcd}'.

\vbox{
\vspace{9cm}
%\begin{figure*}[h]
%\vspace{8cm}
\begin{center}
{{\bf Figure 15.} \it Fit to all electroweak
data in the $S-T$ plane.  Adapted from\cite{PT}}.
\end{center}}
%\end{figure*}

To include fermion masses, one needs to introduce some communication between
the ordinary fermions
(quarks and leptons), $f$, and the condensing fermions, $F$.
The original idea put forth by Dimopoulos and Susskind and by Eichten and
Lane\cite{DSEL} to generate fermion masses in this
context was to introduce {\bf
another} underlying theory---Extended Technicolor (ETC)---whose gauge bosons
connect both the $f$ and $F$ fermions.  Spontaneous breakdown of ETC at a large
scale then produces an effective interaction of the form
\[
{\cal{L}}_{eff} \simeq \frac{1}{\Lambda^2_{ETC}}
(\bar F_L F_R) \bar f_R f_L \, .
\]
\noindent In the presence of the above interaction, the technicolor condensate
$\langle\bar F_L F_R\rangle \sim v_F^3$ gives a mass to the ordinary fermions,
$f$, of order
\[
m_f\simeq \frac{\bigl<\bar F_L F_R\bigr>}{\Lambda^2_{ETC}}
\sim \frac{v_F^3}{\Lambda^2_{ETC}} \, .
\]

\noindent If this formula is to give mass for the top, it is clear that
$\Lambda_{ETC}$ cannot be too large, implying that $\Lambda_{ETC}\sim (1-10)$
TeV.  However, this leads to unacceptably large flavor changing neutral
currents ({\sc fcnc}), since the {\sc etc} interaction also relates ordinary
fermions to ordinary fermions.  For instance, through {\sc etc} interactions
one can get an effective $(\bar ds)^2$ interaction which could directly give
$K-\bar K$ mixing of order
\[
{\cal{L}}^{eff}_{K-\bar K} \sim
\frac{(\cos\theta_c\sin\theta_c)^2}
{\Lambda^2_{ETC}} \,\, (\bar ds)^2 \, .
\]
\noindent The experimental absence of such terms requires $\Lambda_{ETC}
> 100$ TeV.\cite{DE}, leading to a conundrum!

To get round this problem, the concept of
'walking technicolor' was invented.\cite{Holdom}
There is, in fact, a difference between the Fermi scale, $v_F$ (related to
$M_W$), and the condensate scale $\bigl<\bar FF\bigr>$ (which gives $m_f$).
Both are
related to the technifermion self-energy, but they probe different scales in
this function.  The Fermi scale is computed through the contribution of the
technifermion bubble, shown in Fig. 16a, to the $SU(2)$ current correlation
function, while the condensate is proportional to the technifermion self-energy
loop of Fig. 16b.  Thus one has

\begin{eqnarray*}
v_{\rm F}^2  & \sim & \int\frac{d^4 p}{(p^2)^2}
\Sigma^2(p^2) \sim \ \Sigma^2(0)\\[5pt]
\bigl<\overline{F}F\bigr> & \sim &
\int^{\Lambda_{\rm ETC}}\frac{d^4 p}{p^2}\Sigma^2(p^2)
 \sim \ \Lambda^2_{ETC}\Sigma(\Lambda^2_{ETC}) \, . \\
\end{eqnarray*}

\noindent
The approximate dependences given in the above equations ignore all
logarithmic factors.  Furthermore, in the case of the technifermion condensate,
the integral is effectively cut off by the largest scale in the problem where
new physics enters, which here is
$\Lambda_{\rm ETC}$.

\vbox{
\vspace{9cm}
%\begin{figure*}[h]
%\vspace{8cm}
\begin{center}
{{\bf Figure 16.} \it (a) Graph giving $v_F^2$;
(b) graph determining the $<\bar FF>$ condensate.}
\end{center}}
%\end{figure*}

In Yang Mills theories with broken global chiral symmetries, one can
show that\cite{LP}, for large $q^2$
\[
\Sigma(q^2) \ \sim \ \frac{\Sigma^3(0)}{q^2}
\ \sim \ \frac{\Lambda^3}{q^2},
\]
where $\Lambda$ is the dynamical scale of theory.  Clearly, therefore, if
by the time $q^2 \ \sim \ \Lambda^2_{\rm ETC}$,
$\Sigma(q^2)$ has already reached its asymptotic value, then
\[
v_{\rm F} \ \sim \ \Sigma(0) \ \sim \ \Lambda_{\rm TC}
\quad\mbox{and}\quad
\big<\overline{F}F\big> \ \sim \ \Lambda_{\rm ETC}^2
\Sigma(\Lambda_{\rm ETC}^2) \ \sim \ %
\Lambda_{\rm ETC}^2\frac{\Lambda_{\rm TC}^3}{\Lambda_{\rm ETC}^2}
\ \sim \ \Lambda_{\rm TC}^3;
\]
whence
$\left\langle\overline{F}F\right\rangle \sim v_{\rm F}^3$.
However, if $\Sigma(q^2)$ evolves slowly towards
its asymptotic value (`walking technicolor')
it could well be that
\[
\Sigma(\Lambda_{\rm ETC}^2) \gg
\frac{\Lambda_{\rm TC}^3}{\Lambda_{\rm ETC}^2}.
\]
 Then\cite{Holdom}
\[
\bigl<\overline{F}F\bigr> \sim \Lambda_{\rm ETC}^2
\Sigma(\Lambda_{\rm ETC}^2) \gg
\Lambda_{\rm TC}^3 \sim v_{\rm F}^3.
\]
\noindent This behavior is sketched in Fig. 17 below.

\vbox{
\vspace{8cm}
%\begin{figure*}[h]
%\vspace{8cm}
\begin{center}
{{\bf Figure 17.} \it Behavior of the
technifermion self-energy in a walking technicolor theory.}
\end{center}}
%\end{figure*}

In theories where the above behavior holds, one
can get large masses,
having relatively large $\Lambda_{\rm ETC}$,
from a technicolor mechanism
\[
m_f \sim \frac{\big<\overline{F}F\big>}
{\Lambda_{\rm ETC}^2}.
\]
However now, because $\Lambda_{\rm ETC}$ is large, {\sc fcnc} will be
suppressed.
One can construct walking technicolor ({\sc wtc}) models by having
enough technifermions so that the {\sc wtc} coupling
is almost not asymptotically free.\cite{WTC}  In effect,
this means that the {\sc etc} dynamics is not really
 decoupled from the {\sc wtc} dynamics.  As a result, the
`scaled up {\sc qcd}' derivation of $S \sim +O(1)$
no longer holds.\cite{SHL}
Furthermore, in semi-realistic {\sc wtc} models one makes use of
this non-decoupling
of {\sc wtc} and {\sc etc} dynamics to generate
fermions mass splittings, through the different dependence of $\Sigma_u$ and
$\Sigma_d$ on $\Lambda_{\rm ETC}$.\cite{H2}

Once one admits that the {\sc wtc} and the {\sc etc} dynamics are coupled--and
one introduces sufficient complexity in these theories to try to make them
realistic--one has raised the dynamical ante to such an extent that it may be
worthwhile considering other possible options.  In my view, it appears more
sensible instead of introducing an {\sc etc} theory to connect the ordinary
fermions, $f$, to the technifermions, $F$, to appeal to compositeness to make
this connection.  If the $f$ and $F$ states are made up of common
constituents--preons--then the preon dynamics will give rise to effective
4-fermion interactions which play an analogous role to {\sc etc} interactions,
with the dynamical scale of the preon theory $\Lambda_c$ substituting for
$\Lambda_{\rm ETC}$.

I would like to briefly illustrate how this works by discussing a semirealistic
model which I developed in collaboration with S. Khlebnikov.\cite{KP}  The
model is based on 3 replicas of an $SU(6)$ chiral preon theory, augmented by
additional vector-like interactions which are felt by some of the preons
but not
by others.  The model gives one generation of quarks (say $t$ and $b$) which
dynamically have $m_t \gg m_b$\footnote{The model has no $U(1)$ anomalies, even
in the absence of leptons, due to the presence of technifermions with
appropriate $U(1)$ charges.}.  Two of the $SU(6)$ preon theories have 10 preons
in the fundamental representation and 1 preon in the symmetric adjoint
representation, while the third preon theory has a doubling of these degrees of
freedom.  The first two theories will give rise among their bound states to
$t_R$ and $b_R$, while the last preon theory will produce the
$
\left(\begin{array}{l}
t \\ b \\
\end{array}\right)_L
$
\noindent
doublet.  The preons in this model---besides
feeling the $SU(2)\times U(1)$ interactions
also feel three vector interactions:
color, technicolor and metacolor.  The
assignments of the preons under these groups are shown below:

\begin{eqnarray*}
\parbox{4em}{SU(2) singlets}
& \left\{
\begin{array}{rcl}
%\begin{minipage}[c]{16em}
%\vspace*{-15pt}
%\begin{eqnarray*}
\mbox{SU(6)}_{t_{\rm R}} & : &
\overline{3}\,\Box_t^c + \overline{3}\,\Box_t^{\rm T}
+ 4\,\Box_t^{\rm M} + \overline{\strut\Box\!\Box}_t\\[2ex]
\mbox{SU(6)}_{b_{\rm R}} & : &
\overline{3}\,\Box_b^c + \overline{3}\,\Box_b^{\rm T}
+ 4\,\Box_b^{\rm M} + \overline{\strut\Box\!\Box}_b
%\end{eqnarray*}
%\vspace*{-18pt}
%\end{minipage}\right.
%\hspace*{6em}
\end{array}
\right.
\\[4ex]
\parbox{4em}{SU(2) doublets}
&\left.
\begin{array}{rcl}
%\begin{minipage}[c]{16em}
%\begin{eqnarray*}
\mbox{SU(6)}_{\rm L} & : & %
\overline{3}\pmatrix{\Box^c\cr\Box^c\cr}_{\rm L}
+ \overline{3}\pmatrix{\Box^{\rm T}\cr\Box^{\rm T}\cr}_{\rm L}
+ 4\,\Box_1^{\rm M} + 4\,\Box_2^{\rm M}
+ \pmatrix{\overline{\strut\Box\!\Box}\cr
\overline{\strut\Box\!\Box}\cr}_{\rm L}.
%\end{eqnarray*}
%\end{minipage}\right.
\end{array}
\right.
\end{eqnarray*}

\noindent
Here $c$ labels the SU(3) color, $T$ is on SU(3) technicolor
and $M$ the SU(4) metacolor. Further, to avoid anomalies, the singlet
preons are right-handed Weyl fermions, while the doublet preons are left-
handed.

If one turns off all the gauge interactions, the preon theories have certain
global chiral symmetries.  One can argue dynamically ---
through anomaly matching and complementarity\cite{KP} ---
that each of the $SU(6)$ preon theories
preserves a global chiral symmetry
\[
H=SU(6)\times SU(4)\times U(1)\, ,
\]
which is somewhat smaller than the
nominal chiral symmetry of the theory.  The presence
of this chiral symmetry gives rise to certain massless bound states which have
the quantum numbers of the preon composite
\[
\Box \, \Box \,
\overline{\strut\Box\!\Box}
\]
Only states involving two preons with $SU(6)$ quantum numbers, or an $SU(6)$
and an $SU(4)$ quantum number bind.  If one identifies the $SU(4)$ with
metacolor, then the metacolor singlet states--which will be denoted by
$B$-transform under the 15-dimensional representation of $SU(6)$.  The other
states--denoted by $B^\prime$-transform according to the fundamental
representation of the $SU(6)$ and $SU(4)$.  Thus the bound states which are
massless in the theory are:
\[
B \sim (15,1) \, ;  \qquad  B^\prime \sim (6,4) \, .
\]
Since the preons transforming according to the global $SU(6)$ are assigned
either to a $\overline{3}$ of color or a $\overline{3}$ of technicolor, one
sees that the $B$ bound states contain the desired quarks,
as well as some other
states carrying technicolor quantum numbers.  Under $SU(3)_c\times SU(3)_T$
one has
\[
B=(3,1)+(1,3)+(\overline{3},\overline{3}) \, .
\]

Note that since we have 3 distinct preon theories one will find 3 different $B$
and $B^\prime$ massless bound states, which will include, respectively, a
$t_R$, a $b_R$ and a
$
\left(\begin{array}{l}
t \\ b \\
\end{array} \right)_L \, .
$
When one now includes the gauged vectorial interactions not all of these
states will be relevant.  For instance, once the metacolor $SU(4)$ is gauged,
the $B^\prime$ states will bind into metahadrons with masses of order
$M\sim\Lambda_4$--with $\Lambda_4$ being the metacolor dynamical scale.
Furthermore if, as in QCD, these metacolor
interactions cause the formation of
condensates, these condensates will tie the left and right sectors of the model
together.  Indeed, instead of having a separate $SU(6)$ global for each preon
theory, these condensates now preserve
only a common diagonal $SU(6)$\cite{KP}.

\vbox{
\vspace{7cm}
%\begin{figure*}[h]
%\vspace{6cm}
\noindent
{\bf Figure 18.} {\it Formation of an effective {\sc etc} interaction
as a result of the metacolor condensates tying together the effective
interactions caused by the common preon constituents of
$B_L,B_L^\prime$ and $B_R,B_R^\prime.$}}
%\end{figure*}

The formation of the metacolor condensates, in conjunction with the effective
4-fermion interactions arising from each preon theory which tie $B$ and
$B^\prime$ bound states of the same helicity together, give rise to additional
4-Fermi interactions of the {\sc etc} type.  This is shown schematically in
Fig. 18.
The effective {\sc etc} scale here,
however, is different depending on whether
the $B_R$ bound states arise from the $SU(6)_{t_R}$ or the $SU(6)_{b_R}$ preon
theory:
\[
\Lambda_{eff}^t \simeq \biggl(\frac{\Lambda_L}{\Lambda_4}\biggr)
\Lambda_{Rt} \, ; \qquad
\Lambda_{eff}^b \simeq \biggl(\frac{\Lambda_L}{\Lambda_4}\biggr)
\Lambda_{Rb} \, .
\]
Since the $B$-states contain both quarks and techniquarks, these effective {\sc
etc} interactions will give mass to the quarks after one turns on the
technicolor gauge interactions.  One finds in this way

\begin{eqnarray*}
m_t & \simeq & \left[\frac{\Lambda_4^2\,\Lambda_{\rm TC}^3}
{\Lambda_{\rm L}^2\,\Lambda_{{\rm R}t}^2}\right]
 \ = \ K\big<r_t^2\big>\\[2ex]
m_b & \simeq & \left[\frac{\Lambda_4^2\,\Lambda_{\rm TC}^3}
{\Lambda_{\rm L}^2\,\Lambda_{{\rm R}b}^2}\right]
 \ = \ K\big<r_b^2\big>
\end{eqnarray*}

\noindent
That is, here the masses of the $t$ and $b$ quark are different
since they have different sizes.
However, these states are both almost elementary, since
mass generation occurs at scales much larger
than the binding scale.
This is shown schematically in Fig. 19.

\vbox{
\vspace{9cm}
%\begin{figure*}[h]
%\vspace{8cm}
\noindent
{\bf Figure 19.} {\it Scales giving rise to the top and bottom mass
hierarchy in the model: $\bigl<r_C\bigr>\sim 1/\Lambda_{Ri}$---
binding scale; $\bigl<r_4\bigr>\sim 1/\Lambda_4$---ties left
$\leftrightarrow$ right; $\bigl<r_T\bigr> \sim 1/\Lambda_{TC}$
---breaks {\sc ew} symmetry.}}
%\end{figure*}

It is easy mechanically to construct family replications in these
type of models, obtaining `by hand' different masses
for quarks and leptons. However, these models are unsatisfactory
because:\cite{KP}
\begin{enumerate}
\item[i.)] they give no real understanding of the peculiar
hierarchy seen, except\linebreak
mass $\sim$ (size)$^2$;

\item[ii.)] the models have severe technical difficulties,
since they have many preons and the vectorial gauge
interactions are not asymptotically free;

\item[iii.)] the models also have phenomenological difficulties,
since they in general have remnant family symmetries
leading to a trivial Kobayashi-Maskawa matrix.
\[
V_{\rm CKM} \ = \ 1\!\!1\,.
\]
\end{enumerate}

%%%%%%%%%%%%%%%%%%%%%%%%%
%%%%%%%%%%%%%%% SECTION 4
%%%%%%%%%%%%%%%%%%%%%%%%%
%\input{partfour.tex}
%> $.PECCEI.partfour
%%%%%%%%%%LECTURE 4

\section{The question of forces}

Although interactions in the {\sc sm} are totally
defined by specifying the {\sc sm} gauge group
\[
\mbox{SU(3)}\times\mbox{SU(2)}\times\mbox{U(1)}
\]
 there are a number of open questions which are left unanswered.
Some of these questions are:
\begin{enumerate}

\item Why are the strong and electroweak forces described by the
groups SU(3) and SU(2)$\times$U(1), respectively?
\footnote{Theoretically {\sc qcd} = SU(5) is just as reasonable.
Only experiment tells us that the number of colors is 3.}

\item Is there any relation between gravity---the only other force
we know---and the {\sc sm} forces?

\item Are there any other forces in nature besides those of the
{\sc sm} and gravity?

\item Does the remarkable near unification of the {\sc sm} coupling
constants at high energy we discussed earlier
have more than accidental significance?
\end{enumerate}

Let us focus for the moment on this last point.
The apparent coupling constant unification at high scales
observed could be due
to the existence of a Grand Unified Theory ({\sc gut}) which suffers
spontaneous breakdown of the unifying gauge group G:
\[
\rm G \to SU(3)\times SU(2)\times U(1)
\]
at the unification scale $M_X$\cite{GG}\cite{GQW}.  This behavior is
shown schematically in Fig. 20.
%\pagebreak

\vbox{
\vspace{9cm}
%\begin{figure*}[h]
%\vspace{8cm}
\begin{center}
{\bf Figure 20.}  {\it Schematic plot of the evolution of coupling
constants in a {\sc gut} spontaneously broken at the
scale $M_X$.}
\end{center}}
%\end{figure*}

Let me describe briefly some other nice feature of {\sc guts},
besides coupling constant unification.
If G is not a product group, then
the idea of grand unification can
explain charge quantization, since
$Q_{\rm em}$ becomes one of the generators
of $G$.  Thus in {\sc gut}s the quark charges
and the lepton charges are naturally related.
Indeed, the known quarks and leptons fit very nicely into
representations of simple candidate {\sc gut} groups
\[
\overline{5} \ \oplus \ 10 \subset \mbox{SU(5)} \, ;
\qquad 16 \subset \mbox{SO(10)} \, .
\]

\noindent
It is useful to classify the fermions in terms of
left-handed fields only, using \\
$\Psi_{\rm L}^c = C(\overline{\Psi}_{\rm R})^T$.
Then for SU(5),\cite{GG} since $Q_{em}$ is a generator, the
$\overline{5}$ must contain $d^c$ and the lepton
doublet:\footnote{Note that
$\overline{5} \supset \overline{2}$ of SU(2) and
$C\pmatrix{\nu\cr e\cr} = \pmatrix{e\cr -\nu\cr}$.}
\[
\overline{5} \ : \ \Psi_{\rm L}^a \sim
\pmatrix{d^c\cr d^c\cr d^c\cr e\cr -\nu\cr}_{\rm L}  \, ,
\]
while the 10 contains:
\[
10 = \Psi_{ab\rm L} \sim \frac{1}{\sqrt{2}}
\pmatrix{
0&u^c&-u^c&-u&-d\cr
-u^c&0&u^c&-u&-d\cr
u^c&-u^c&0&-u&-d\cr
u&u&u&0&-e^c\cr
d&d&d&e^c&0\cr} \, .
\]

The group SU(5) has 24 generators.  Of these
12 are associated with the generators of
SU(3)$\times$SU(2)$\times$U(1) and 12 are additional.
Associated with these additional generators are 12
superheavy gauge bosons $X$, $Y$ with masses of
order $(M_X)$, resulting from the spontaneous breakdown
\[
\mbox{SU(5)} \to \mbox{SU(3)}\times\mbox{SU(2)}\times\mbox{U(1)}.
\]
 Schematically, the gauge fields can be organised
into a 5$\times$5 matrix

\begin{center}
$
\left(\begin{array}{ccc|cc}
& & & & \\
\multicolumn{3}{c|}{A_3 - \sfrac{2}{\sqrt{15}}B}
& \overline{X} & \overline{Y} \\
& & & & \\
\hline
& X & & & \\
& & & \multicolumn{2}{c}{
W + \frac{3}{\sqrt{5}}B}\\
& Y & & & \\
\end{array}\right)
$
\end{center}

\noindent
where $A_3$ corresponds to the SU(3) bosons,
$X$ and $B$ are the heavy bosons
and $W,B$ corresponds to SU(2) and U(1) bosons,
respectively.

In SU(5) the electromagnetic charge is
\[
Q_{\rm em} = I_{23} + \sqrt{\sfrac{5}{3}}\,I_{24},
\]
This can be readily checked since,
in the fundamental representation,
these diagonal generators
(normalized to $\mbox{Tr} \lambda_a \lambda_b=\delta_{ab}$)
are given by:
\[
\frac{\lambda_{23}}{2} \ = \ \pmatrix{
0&&&&\cr
&0&&&\cr
&&0&&\cr
&&&1/2&\cr
&&&&-1/2\cr}\qquad
\frac{\lambda_{24}}{2} \ = \ \frac{1}{\sqrt{15}}\pmatrix{
-1&&&&\cr
&-1&&&\cr
&&-1&&\cr
&&&3/2&\cr
&&&&3/2\cr}.
\]
Comparing with the SU(2)$\times$U(1) definition
$Q_{em} = T_3 + B$ we see that what unifies with $g_2$ and $g_3$
is not $g_1$,
but $\sqrt{\sfrac{5}{3}}g_1$.
Thus for an SU(5) {\sc gut} one expects
\[
g_0(M_X^2) = \sqrt{\sfrac{5}{3}}g_1(M_X^2)
\ = \ g_2(M_X^2) \ = \ g_3(M_X^2),
\]
where $g_0(M_X^2)$ is the SU(5) coupling at
the breaking scale, $M_X$.  This relationship fixes
the weak mixing angle at the {\sc gut} scale:\cite{CG}
\[
\sin^2\theta_W(M_X) \equiv \frac{g_1^2(M_X^2)}
{g_2^2(M_X^2) + g_1^2(M_X^2)} = \frac{3}{8} \, .
\]
 Running this down to $M_Z$, using as a unification
scale $M_X = 10^{16}\,$GeV and the 1-loop RGE, gives
\[
\sin^2\theta_W(M_Z) = \left\{\begin{array}{ll}
0.194 & \mbox{\sc gut}\\[1ex]
0.235\;  & \mbox{\sc susy gut}\\
\end{array}\right.
\]
to be compared with the experimental value of
$\sin^2\theta_W(M_Z) = 0.2260 \pm 0.0024$.\cite{sine}
A more refined calculation\cite{PL} actually improves
the {\sc susy gut} value to below 0.230, whereas the {\sc gut} with
ordinary matter has problems still.
The {\sc gut} value can be improved to
$\sin^2\theta_W(M_Z)=0.212$ by using
\mbox{$M_X = 10^{14}\,$GeV}
but this leads to other difficulties, predicting too high a
rate for proton decay.  Clearly the above analysis is the converse
of the extrapolation from low energy to high energy discussed
earlier and it shows that unification with {\sc susy}
works, but otherwise it doesn't--at least naively.

Let me discuss the issue of proton decay a bit further.
Since leptons and quarks are in the same SU(5)
representations, $X$ or $Y$ boson exchange can change
quarks into leptons, mediating proton decay.
Fig. 21 shows an example of a diagram leading to
proton instability, as a result of $X$ exchange:
%\pagebreak

\vbox{
\vspace{9cm}
%\begin{figure*}[h]
%\vspace{8cm}
\begin{center}
{{\bf Figure 21.} {\sc gut} \it transitions leading to proton decay.}
\end{center}}
%\end{figure*}

\noindent
Since $M_X$ is very heavy, one gets an effective,
4-fermion Lagrangian for this transi\-tion\cite{EGN}
\[
{\cal L}_{\rm eff} = \sqrt{2}\pi
\frac{\alpha_u(M_X^2)}{M_X^2}\left\{
\varepsilon_{abc}(\overline{u}_{c\rm L}^c\gamma^\mu u_{b\rm L})
\times(2\overline{e}_{\rm L}^c\gamma^\mu\,d_{a\rm L}
+ \overline{e}_{\rm R}^c\gamma^\mu\,d_{a\rm R})\right\}
\]
where $\alpha_u(M_X^2)$ is the grand unified coupling.
The above Lagrangian gives a decay rate proportional to $M_X^{-4}$,
or a proton lifetime for the dominant SU(5) $p\to e^+\pi^0$
mode\cite{PL}.
\[
\tau(p \to e^+\pi^0) \ = \ 10^{31 \pm 0.7}
\left\{\frac{M_X({\rm GeV})}{4.6\times 10^{14}}
\right\}^4 \mbox{years},
\]
 which is in conflict with the experimental bound\cite{Kamioka}
 \[
\tau(p \to e^+\pi^0)_{\rm exp} > 10^{33}\,\mbox{years}\, ,
\]
\noindent
if $M_X=10^{14}$ GeV.

{\sc susy gut}s fare better in this respect,
as $M_X \sim 10^{16}\,$GeV gives
\[
\tau(p \to e^+\pi^0) \ = \ (10^{35}\mbox{--}10^{36})\mbox{ years.}
\]

\noindent
However, in these theories, one must be careful since
there are other contributions to proton decay besides
$X$ and $Y$ exchange.\cite{dim5}  In particular,
there is another process in {\sc susy gut}s
involving neutralinos,
$\widetilde{\delta} = \{\widetilde{\gamma}, \widetilde{Z},
\widetilde{H}_1, \widetilde{H}_2\}$
and a colored Higgsino, $\widetilde{H}_3$,
whose core really involves a dimension 5
operator containing two fermions and two sfermions.
The box diagram involving this dimension 5 operator is
shown in Fig. 22.

\vbox{
\vspace{7cm}
%\begin{figure*}[h]
%\vspace{6cm}
\begin{center}
{{\bf Figure 22.} \it Dangerous dimension 5 graph contributing
to proton decay in {\sc susy gut}s.}
\end{center}}
%\end{figure*}

Due to this dimension 5 operator, the proton lifetime is
\[
\tau[{\rm dim\ 5}] \ \sim \ \frac{1}{h^4}
\frac{1}{\alpha_U^2}\left[ M_{\widetilde{H}_3}\,
\widetilde{m}_{\rm eff}\right]^2
\]
where the $1/h^4$ comes from the Yukawa couplings.
Because of the appearance of the neutralino mass $\widetilde{m}_{eff}$,
even taking $M_{\widetilde H_3}\sim M_X$ gives a
lifetime
only proportional to $M_X^2$,
not $M_X^4$.  To escape troubles with experiment,
this requires both $M_{\widetilde{H}_3}$ to be big,
$M_{\widetilde{H}_3} \gg M_X$
and the {\sc susy} breaking parameter,
$\widetilde{m}_{\rm eff}$ not to be too light.
The expected range is very uncertain\cite{AN}
\[
\tau[{\rm dim\ 5}] \ \simeq \ 10^{29 \pm 4}\,\mbox{years},
\]
with the dominant decay modes being
\[
p \to K^+\,\nu_\tau \qquad\mbox{and}\qquad
p \to K^0\,\mu^+\,.
\]

Since the present limits for these modes
are already near $10^{32}$ years,\cite{Kamioka}
there is some interest in {\sc susy gut}s models with essentially
no dimension 5 operators, such as flipped SU(5)\cite{Flip}.
Flipped SU(5) is not a true {\sc gut},
since the gauge group is SU(5)$\times$U(1),
but the extra U(1) allows flipping the assignments
of quarks and leptons: $\nu \leftrightarrow e$
and  $u \leftrightarrow d$.  For instance, one has in this case
\[
\overline{5} \ = \ \pmatrix{
u^c\cr u^c\cr u^c\cr \nu\cr e\cr}_{\rm L}
\qquad 1 \ = \ e_{\rm L}^c.
\]
Flipped SU(5) arises quite naturally in various
superstring models.
Thus, this group is of some intrinsic interest
anyway.  What makes it more interesting in this
context is that, as a result of the flipped
assignments, what sits in the 10 of flipped SU(5)
is not $e_L^c$ but $\nu_L^c$.  Thus in flipped SU(5),
one can break the {\sc gut} group to the {\sc sm}
using Higgses which are in the 10-dimensional
representation of SU(5), because a scalar with the
quantum numbers of $\nu_L^c$ has a {\sc vev} which
preserves the {\sc sm} group.  The Higgs triplet of
Fig. 22 in flipped SU(5) can get mass from this {\sc vev}
and one can well imagine that
\[
\bigl<10_{\mbox{Higgs}}\bigr> \gg M_X
\]
leading to a large mass for $M_{\widetilde H_3}$.\footnote{The
color triplet Higgsinos in the {\sc susy} SU(5) extension of
the {\sc sm} can get their mass via a superpotential term
coupling two $10_{\mbox{Higgs}}$ with a $\overline{5}_{\mbox{Higgs}}$.}

I want to, finally, comment briefly about another possible
{\sc gut}: SO(10).\cite{FMG}
Perhaps the nicest feature of SO(10) as a
{\sc gut} is that all fermions
sit in a single representation of
this group, the 16 dimensional representation.
The 16 of SO(10) decomposes in terms of SU(5) as:
\[
16  \longrightarrow  10 \ + \ \overline{5} \ + \ 1
\]
The SU(5) singlet here corresponds to
$\nu_{\rm L}^c$.
Because SO(10) is a rank 5 group, it can be broken
to the rank 4 {\sc sm} either directly or in two steps
(sequential breaking).
In the first case there is only one mass scale
\begin{eqnarray*}
\mbox{SO(10)} & \longrightarrow &
\mbox{SU(3)$\times$SU(2)$\times$U(1)}\\
& M_X &
\end{eqnarray*}
 and one finds the same difficulties with
$\sin^2\theta_{\rm W}$ and proton decay, in the
absence of {\sc susy}, as with an SU(5) {\sc gut}.
In the second case, there are two mass scales and
the {\sc gut} breaking scale can be at $M_X = 10^{16}$
GeV, leading to perfectly consistent {\sc gut}s without
{\sc susy}.  The chains
\begin{eqnarray*}
\mbox{SO(10)} & \longrightarrow &
\mbox{SU(2)$_{\rm L}\times$SU(2)$_{\rm R}\times$SU(4)}
\ \longrightarrow \ \mbox{SU(3)$\times$SU(2)$\times$U(1)}\\
& M_X & \!\!\!= 10^{16}\mbox{GeV}
\hspace*{6em} M_I = 10^{12}\mbox{GeV}
\end{eqnarray*}
 or
\begin{eqnarray*}
\mbox{SO(10)} & \longrightarrow &
\mbox{SU(2)$_{\rm L}\times$SU(2)$_{\rm R}$}
+ \mbox{U(1)$_{\rm B-L}\times$SU(3)}
\ \longrightarrow \ \mbox{SU(3)$\times$SU(2)$\times$U(1)}\\
& M_X & \!\!\!= 10^{16}\mbox{GeV}
\hspace*{8em} M_I = 10^{10}\mbox{GeV}
\end{eqnarray*}
\noindent
give measurable values for the proton lifetime and
$\sin^2\theta_W$ in agreement with experiment, but
are not very predictive otherwise.
The presence of an intermediate mass scale has
beneficial features related to neutrino masses, since it
gives a natural mechanism, the see-saw mechanism,\cite{SS}
for explaining why $m_\nu \ll m_e$ (see below).
However, it also introduces another hierarchy problem,
why is $M_X \gg M_I \gg v_{\rm F}$?

It is difficult to be fully convinced of {\sc gut}s
until the `smoking gun' of proton decay is found.
There may be some hope here, since
the new SuperKamiokande detector, being built
now in Japan, will push
the limits on the proton lifetime to
$\tau_p \sim 10^{33}$--$10^{34}\,$years.
Nevertheless, {\sc gut}s do possess enough attractive
features to make their continued investigation very worthwhile.
One of these features is connected with the matter-antimatter
asymmetry in the universe discussed in Kolb's lectures.
The production of a matter-antimatter asymmetry
in the early universe requires that baryon number
violating processes should go out of equilibrium
in the early universe, and this can happen due to
B violating decays of the
heavy $X$ and $H_3$ states in {\sc gut}s.
It remains to be seen, however, whether
the baryon to photon ratio
\mbox{$n_B/n_\gamma \approx 10^{-10}$}
originates from $B-L$ violating processes at the
{\sc gut} scale (we owe our existence to {\sc gut}s).
or it
originates from $B+L$ violating
processes at the electroweak scale
(we owe our existence to the {\sc sm}).

%%%%%%%%%%%%%%%%%%%%%%%%%
%%%%%%%%%%%%%%% SECTION 5
%%%%%%%%%%%%%%%%%%%%%%%%%
%\input{partfive.tex}
%> $.PECCEI.partfive

%%%%%%%%%%%%%%%%%%%LECTURE 5

\section{The question of matter}

Perhaps the hardest open problems in the {\sc sm} are related to the question
of matter.   Three of these questions are particularly striking:
\begin{description}
\item[\rm(i)] Why do we see the fermion representations we see in nature?

\item[\rm(ii)] Why are the fermions that make up the matter we see chiral?

\item[\rm(iii)] Why do quarks and leptons appear in repetitive family
       structures and why, apparently, are there only three families?
\end{description}
In my opinion (ii) is probably answered by Georgi's survival
hypothesis,\cite{Georgi}
but the other two are very deep.  In fact, ultimately if we really
understood (i) and (iii) I believe we could calculate fermion masses
and mixings.

Georgi's survival hypothesis can be succinctly summarized as follows:
fermions, unless prohibited by their chiral assignments, will get
a mass consistent with the largest scale in the theory.
Thus chiral fermions are the only ones which are light.  All the examples
we discussed are in accord with this hypothesis.  Certainly preon
models are built this way, since chiral fermions are the ones
which are light.
This is also true in {\sc gut}s.  For instance in
SO(10) fermions are in the 16 representation only.
If they were both in the 16 and in the $\overline{16}$,
one could write an invariant mass term
\[
{\cal L}_{\rm mass} = M\,\overline{16}\cdot 16.
\]
and there would be no reason for $M$ not to be of $O(M_X)$.

Note that right-handed neutrinos are not in a chiral representation
of the {\sc sm}, since $\nu_{\rm R} \sim 1$.
Thus they are allowed to have a large Majorana mass term
\[
 {\cal L}_{\rm mass}  = m_{\rm R}
(\nu_{\rm R}^T\,C\,\nu_{\rm R}
+ \overline{\nu}_{\rm R}^T\,C\,\overline{\nu}_{\rm R}).
\]
Such a term could explain why neutrinos are light
via the see-saw mechanism,\cite{SS}
since including Dirac masses (of order $m_\ell$)
gives a neutrino mass matrix with vastly different eigenvalues:
\[
\left(
\begin{array}{cc}
0 & m_{\rm D}\\
m_{\rm D} &
m_{\rm R}
\end{array}
\right)
\quad\Longleftrightarrow\quad
%\begin{minipage}{15em}
\begin{array}{l}
(m_\nu)_{\mbox{light}}  \simeq  m_{\mbox{D}}^2/m_{\mbox{R}}
\ll m_{\mbox{D}} \\
(m_\nu)_{\mbox{heavy}}  \simeq  m_{\mbox{R}} \, .
\end{array}
%\end{minipage}
\]

As we saw, {\sc gut}s provide a concise way to characterize the fermion matter
we know; for instance, by classifying the quarks and leptons in the
$\overline{5}$ and 10 representations of SU(5) or the 16 of SO(10).  However
{\sc gut}s do not fix the representation content of the fermions.\footnote
{There
is a mild restriction on which representations are allowed coming from the
requirement that there should be no gauge anomalies.  In SU(5) having a
$\overline{5}$ and 10 dimensional representation guarantees this to be the
case.  However, a matter content made up of four $\overline{5}$ and a 15
would also be anomaly free, but this is not nature!}  Nevertheless,
some progress in understanding fermion mass patterns emerges from
{\sc gut}s, since the larger symmetry can restrict the form
of the Yukawa couplings, if only a limited number
of Higgs representations are included.  In the case of
SU(5), with a $\overline{5}$ of Higgs particles, $H_1$, plus
another 5 of Higgs particles, $H_2$, one has only two
distinct Yukawa couplings
\[
{\cal L}_{\rm Yukawa} = \Gamma_1\,\overline{5}_f\,10_f\,\overline{5}
+ \Gamma_2\,10_f\,10_f\,5\,.
\]
Then $\left<H_1\right> \neq 0$ gives a common mass
to the leptons and to the charge $-1/3$~quarks,
while $\left<H_2\right> \neq 0$  gives a mass to the charge $+2/3$~quarks.

The equality of the lepton and $-1/3$~quark
mass matrices at the {\sc gut} scale implies relations
between masses at low energy via the {\sc rge}\cite{CEG}
\[
\frac{m_{\rm D}(q^2)}{m_{\rm L}(q^2)} =
\frac{m_{\rm D}(M_X^2)}{m_{\rm L}(M_X^2)}_{}
\left(\frac{\alpha_3(q^2)}{\alpha_3(M_X^2)}\right)^{4/b_3}
\]
In SU(5) with just $H_1 = \overline{5}$
as a Higgs, the {\sc gut} scale ratio is unity.  In this
case the above equation
gives a good prediction of
$m_b/m_\tau$ in ordinary SU(5) (where $b_3 = 7$)\cite{CEG}.
\[
m_b/m_\tau \simeq 2.8 \, ,
\]
 but not such a good prediction in {\sc susy} SU(5),
since here $b_3 = 3$ gives $m_b/m_\tau \simeq 4$.
However, incorporating two-loop effects one can
obtain agreement also for {\sc susy} SU(5) provided a
large region in
$\tan\beta = v_2/v_1$ is excluded\cite{PLO}.  I will
return to this point below.
However, other predictions are not so good.
For example, the equality of the lepton and
charge -1/3 matrices at the {\sc gut} scale predicts
also that
\[
\frac{m_s}{m_d} \ = \ \frac{m_\mu}{m_e} \, .
\]
But the {\sc lhs} is around 20 from  current algebra
estimates\cite {GL}, while the {\sc rhs} is 206 from experiment.

This circumstance has generated a lot of interest in having a
better set of {\sc gut} mass matrix relations.
A representative example is the work of Dimopoulos,
Hall and Raby\cite{DHR} which uses
mass matrices which do
not quite have $m_{\rm D} = m_{\rm L}$ at $M_X$,
but have a pattern (Georgi-Jarlskog
pattern\cite {GJ}) and a texture of zeros first
suggested by Fritzsch\cite{HF}:
\[
M_{\rm D} = \pmatrix{
0 & F & 0\cr
F & E & 0\cr
0 & 0 & D\cr}\;;\qquad
M_{\rm L} = \pmatrix{
0 & F & 0\cr
F & -3E & 0\cr
0 & 0 & D\cr}\;;\qquad
M_{\rm U} = \pmatrix{
0 & C & 0\cr
C & 0 & B\cr
0 & B & A\cr} \, .
\]
The factor of 3 in the 22 matrix element of $M_{\rm L}$
is the Georgi-Jarlskog factor\cite{GJ}\footnote{This factor
is related to the number of colors.  Instead of a $\overline{5}$,
Georgi and Jarlskog\cite {GJ} use a $\overline{45}$ Higgs,
$H_\gamma^{\alpha\beta}$, with $\{\alpha,\beta,\gamma=1,\cdots 5\}$,
which is antisymmetric in $\alpha$ and $\beta$ and traceless.
What gives mass to $M_{\rm D}$ and $M_{\rm L}$ is
$\big<H_\gamma^{\alpha\beta}\big>=\delta_\gamma^\beta
\lambda_\alpha$ with $\lambda_\alpha$ being the diagonal
elements of a 4 $\times$ 4 matrix.  Since there are 3 colors,
and we do not want the {\sc vev} to break color, it
follows that $\lambda_\alpha=\{1,1,1,-3\}$.} and produces
a different patterm of {\sc gut} mass relations.  Namely, in
this case one finds:
\[
m_b(M_X) = m_\tau(M_X)\,;\qquad
m_s(M_X) = \frac{1}{3}m_\mu(M_X)\,;\qquad
m_d(M_X) = 3\,m_e(M_X)\,,
\]
which fixes correctly the
$m_s/m_d$ to $m_\mu/m_e$ ratio---since this ratio, essentially,
does not evolve.

This pattern follows rather naturally
in an SO(10) {\sc gut}\cite{HRR} as a result
of a Higgs 45 breaking.  However, in an SO(10) {\sc gut}
with a Higgs in the 10 dimensional representation to
provide the principal Yukawa couplings, one would
also expect that $D=A$.  That is, that there be a
relation between $m_t(M_X)$ and $m_b(M_X)$.  Indeed,
since the SO(10) 10 representation contains both
an SU(5) 5 and a $\overline{5}$, it is easy to see that
in SO(10) the two SU(5) Yukawa couplings, $\Gamma_1$
and $\Gamma_2$ just reduce to a single Yukawa coupling
\[
{\cal{L}}_{\mbox{Yukawa}} = \Gamma \, 16_f \, 16_f \, 10 \, .
\]
The relation
\[m_b(M_X)=m_t(M_X)
\]
is, however, difficult to evolve to low energy to obtain
realistic values of $m_t$.  A recent analysis in the
context of {\sc susy gut} theories is shown in Fig. 23.
Only for large $\tan\beta$ is there a line where the
unification of all 3 Yukawa couplings is allowed, while
much more of the $\tan\beta$ plane is allowed by not
unifying also the top quark Yukawa.
%\pagebreak

\vbox{
\vspace{9cm}
%\begin{figure*}[h]
%\vspace{6cm}
\noindent
{\bf Figure 23.}  {\it Allowed values in the
$m_t-\tan\beta$ plane assuming
that some Yukawa couplings unify at $M_X$.  The triple unification
line is shown.  The values from $\tan\beta\simeq 1$ to $\tan
\beta\simeq 40$ are excluded.  The Arnowitt bound\cite{AN}
comes from an analysis of proton decay in these models.  Adapted
from\cite{PLO}.}}
%\end{figure*}

Interesting predictions for mass ratios
and mixing angles follow from the mass matrices and
textures at $M_X$ assumed by
Dimopoulos, Hall and
Raby\cite{DHR}.
Using the freedom
of phase redefinition for the quark and lepton fields, the
number of parameters in these mass matrices is reduced to
seven--6 real parameters and a phase.  Thus the effective
mass matrices at $M_X$ are given by
\[
M_D=\left[
\begin{array}{ccc}
0 & fe^{i\phi} & 0 \\
fe^{-i\phi} & e & 0 \\
0 & 0 & d
\end{array} \right] \, ; \qquad
M_L=\left[
\begin{array}{ccc}
0 & f & 0 \\
f & -3e & 0 \\
0 & 0 & d
\end{array} \right] \, ; \qquad
M_U=\left[
\begin{array}{ccc}
0 & c & 0 \\
c & 0 & b \\
0 & b & a
\end{array} \right] \, .
\]
Because Dimopoulos {\it et al.}\cite{DHR} assume a {\sc susy gut} theory,
in addition to the above parameters, one also has
$\tan\beta$ as a free parameter.  These unknowns in the model are
fixed by using as input the experimental masses and mixing angles
which are best known: $m_e$, $m_\mu$, $m_\tau$\cite{PDG};
$m_c(m_c), \,\, m_b(m_b)$; $m_u/m_d$\cite{GL};
$|V_{us}|, \, |V_{cb}|$\cite{PDG}.  This then allows 6
predictions, one of them being $\sin\beta$.  The results
are:

\begin{eqnarray*}
m_t(m_t) & = & 176~ \mbox{GeV} \left[ \frac{m_b}{4.25 \, \mbox{GeV}}\right]
\left[\frac{m_c}{1.25 \mbox{GeV}}\right]
\left[\frac{0.053}{V_{cb}}\right]^2 \\
m_s(1~ \mbox{GeV}) & = & 156~ \mbox{MeV} \\
m_d(1~ \mbox{GeV}) & = & 6.2~ \mbox{MeV} \\
|V_{ub}|/|V_{cb}| & = & 0.05 \left[\frac{m_u/m_d}{0.6}\right]^{1/2} \,
\left[\frac{1.25~\mbox{GeV}}{m_c}\right]^{1/2} \\
\cos\delta & = & (0.53\pm 0.07)
\left[\frac{0.6}{m_u/m_d}\right]^{1/2}
\left[\frac{m_c}{1.25\mbox{GeV}}\right]^{1/2} \\
& & -0.13 \left[\frac{m_u/m_d}{0.6}\right]^{1/2}
\left[\frac{1.25 \mbox{GeV}}{m_c}\right]^{1/2} \\
\sin\beta & = & 0.95 \left[\frac{4.25 \mbox{GeV}}{m_b}\right]^5
\left[\frac{m_c}{1.25 \mbox{GeV}}\right]
\left[\frac{0.053}{V_{cb}}\right]^2 \\
& & \times\left[ \left[\frac{4.25 \mbox{GeV}}{m_b}\right]^{12}
-0.08\right]^{1/2} \, .
\end{eqnarray*}
These results appear quite reasonable, but one needs real data on
$m_t$, $\delta$ and $\beta$ for a true test.  I note here,
parenthetically, that the value of $V_{cb}$ more consonant
with the latest data on the B lifetime is nearer to 0.043\cite{MW}
than the value used here of 0.053.

Although the above attempts may eventually bear fruit, they
do not really provide a deep answer for the nature of
fermionic matter.  Contrast what we know about fermions
with the situation regarding gauge bosons.
Given G we know immediately that the gauge bosons are in the
adjoint representation; therefore the representation of the forces
is fixed by G.
Why are the fermions in nature in the representations that we see?
One could answer this question if there was a {\sc susy}  where,
instead of having separate chiral and vector multiplets
describing matter and forces,
one unified matter and forces by having only vector multiplets.
Unfortunately, this does not work!
The adjoint representation is a real representation
and always includes chirally paired representations
of fermions upon decomposition.

This problem is nicely
illustrated by the following
example.  Consider a {\sc gut} E$_6$, whose adjoint is a 78.
If we decompose the 78 of E$_6$ into representations of the SO(10)
subgroup, we find
\[
78 \ = \ \underbrace{45 \ \oplus \ 1}_{\dst\rm Real}
\ \oplus \ \underbrace{16 \ \oplus \ %
\overline{16}}_{\dst\rm \shortstack{Chirally\\paired}}
\]
The question arises.  Are there ways in this example
where one can
get rid of one of the chiral pairs  and of the real states?
If one could eliminate the real fermions and suppress the wrong
chirality fermions then one could really claim to have achieved a
unification between forces and matter!

This is, roughly, what happens in superstring theories.
I will close these lectures by briefly indicating how this
remarkable fact comes about.  These speculations connected
to superstrings are, in a sense, in a class of their own.
Indeed, they give really beautiful insights into what may be
the nature of matter and forces.  Other ideas I discussed
here--like preon models--are very much more modest
and mechanical.  For instance, in the composite model we
discussed one got color triplets of quarks by
starting from color antitriplets of preons, using the fact
that $\overline{3} \times \overline{3} \supset 3$.  If
this were to be a {\bf real theory}, one would want to
eventually understand why the preons in nature come in
$\overline{3}$ of color!  These questions are premature
in preon models, but are perfectly sensible ones to ask
{\it ab initio} in superstring theories.

String theories were known, almost from the beginning, to
require extra dimensions of space-time for their physical
consistency\cite{strings}.  Whether these extra dimensions
are actually real or not, however, is not so germane
anymore, as one has been able to formulate these theories
directly in 4-dimensions by associating the extra dimensions
with internal degrees of freedom\cite{4dstrings}.
Nevertheless, to understand how some {\bf unwanted} fermions
are made to disappear in string theories, it is useful to
retain the idea of having these extra dimensions, for it is
through the process of {\bf compactifications} of these
extra dimensions that chiral fermions are most easily seen
to emerge.

It has been known for a long time that physical theories in $\rm D>4$
can give sensible $\rm D=4$ theories when the extra
dimensions compactify. For example,
in the original theory of Kaluza and Klein\cite{KK}
gravity in D=5, when one of the dimensions was
compactified, gave rise to gravity and electromagnetism
in D=4.
Consider then chiral fermions in $\rm D>4$ dimensions.
Chiral fermions obey the massless Dirac equation
\[
\Gamma^\alpha\,D_\alpha\,\Psi \ = \ 0.
\]
 where $\alpha = 1,\ldots $D
and $D_\alpha$ is the covariant derivative
containing background fields (gravity, Yang-Mills
fields, etc.).  It is easy to see formally that,
if the theory
compactifies to 4 dimensions, the erstwhile chiral fermions get
a mass:
\[
(\Gamma^\mu\,D_\mu + \Gamma^a\,D_a)\Psi \ = \ 0.
\]
 The first term in the above is the
4-D Dirac kinetic energy operator,
while the second term, containing the $\rm D-4$~dimensional
contribution ($a = 5,\ldots,$ D), acts precisely as a mass
term.

Although, in general, $\Gamma^a\,D_a$
acts as a mass in four dimensions, this will
not be the case if there are `zero modes'
solutions in the $\rm D-4$ dimensional compactified space
\[
\Gamma^a\,D_a\Psi \ = \ 0.
\]
If the above holds, there will remain chiral fermions in D=4.
This is a constraint equation in the D-4 dimensional
compact space, $K$.
In general, solutions to this constraint equation come in repetitive
patterns, which depend on the geometry of $K$.
Usually the number of solutions of
$\Gamma^a\,D_a\Psi = 0$
is related to a topological index of $K$\cite{WW}.  In this scenario
not only the type of chiral fermions (matter)
but also the number of families is the
result of the geometry involved!

A very similar mechanism to the one just outlined
obtains in string theories.  Before describing a
particular realization, it may be useful to very
briefly summarize
the principle features of string theories\cite{strings}:
\begin{description}
\item{i)} One replaces particles with normal modes of a one-dimensional
`string' ($m=0$ excitations) and then quantizes these excitations.

\item{ii)} The theories are only consistent if specific
gauge groups are attached to the excitations and the
dimension of space-time is $\rm D=10$ for superstrings
(or 26 for ordinary strings).

\item{iii)} The massless spectra contains a spin-2 graviton
and the theories give a consistent and finite quantum
theory of gravity.

\item{iv)} In these theories one replaces the point
interactions of field theory by string interactions
occurring on surfaces.
\end{description}

To illustrate how things could work out,
I shall describe a scenario\cite{CN}
associated with the heterotic string,
where $\rm D=10$ is compactified to $\rm D=4$
on a 6 dimensional Calabi-Yau manifold
(this is only one of many possibilities).
The gauge group associated with this theory is\cite{Gross}
\[
\rm G = E_8\times E_8 \, ,
\]
 where the first E$_8$ corresponds to the real world,
and the second to the `hidden' world (ignored here).
In 10-dimensions, one can have Weyl fermions (chiral
fermions) which obey a Majorana condition and thus
are suitable for transforming according to the
adjoint representation\cite{Slansky}.  For the
heterotic string these
chiral fermions are gauginos ({\sc susy} partners of
the E$_8$ gauge bosons).
The SU(3)$\times$E$_6$ decomposition of the 248
gauginos (248 is the dimension of the adjoint representation
of E$_8$) has a similar pattern of the earlier $E_6$
example:
\begin{eqnarray*}
248 & = & \underbrace{(1,\,78)\ \oplus \ (8,\,1)} \ \oplus \ %
\underbrace{(3,\,27) \ \oplus \ (\overline{3},\,\overline{27})}.\\
& & \mbox{\shortstack{Real\\ representations}}\hspace*{1cm}
\mbox{\shortstack{Complex\\ conjugate\ pairs}}
\end{eqnarray*}
 The 6 dimensional manifold, $K$, where the compactification
is assumed to occur\cite{CN}
is taken to be a Calabi-Yau
manifold which has the special property that the zero modes
of the Dirac equation on $K$ have non-trivial SU(3) properties:
\[
\Gamma^a\,D_a \Psi \ = \ 0
\]
Thus in this case the zero modes are:
$\Psi \sim (3,\,27)$ and $(\overline{3},\,\overline{27})$.
Furthermore, in the compactification
the gauge group is reduced to E$_6$, or one of its subgroups:
\begin{eqnarray*}
\rm E_8 & \longrightarrow & \rm E_6 \, . \\
\mbox{$\rm D=10$} & & \mbox{$\rm D=4$}
\end{eqnarray*}
In terms of $E_6$,
the chiral matter that appears consists of
\[
n_f\cdot 27 + \delta(27 + \overline{27}) \, ,
\]
 with both $n_f$ (the number of families!)
and $\delta$ related to the topological
properties of $K$.  In fact, Candelas {\it et al.}\cite{CN}
show that
$n_f = \frac{1}{2}|\chi|$, where $\chi$ is the
Euler characteristic of $K$.
Note that the matter here is that of the {\sc sm} (or of an SU(5)
{\sc gut}), with
some extra matter degrees of freedom which
are vectorial:
\begin{eqnarray*}
27 & = & \overbrace{\ \underbrace{\overline{5} \ \oplus \ 10}
\ \oplus \ \underbrace{1 \ \oplus \ 5
\ \oplus \ \overline{5} \ \oplus \ 1}\ }^{\dst\rm SU(5)}\\
\uparrow & & \mbox{ chiral}\hspace*{5em}\mbox{vectorial}\\
\mbox{ E$_6\!\!$} & &
\end{eqnarray*}

Two remarks are in order which transcend this simple example:
\begin{description}
\item{i)} In general, for reasonably low levels of the
associated string Kac-Moody algebra\cite{RW}, one cannot
be left with a {\sc gut} group after compactification
because matter scalars do not contain the adjoint
representation.  With scalars in an $E_6$ 78 one
can effect the breaking:
\begin{eqnarray*}
\rm E_6 & \longrightarrow & {\rm SU(3)} \times {\rm SU(2)}
\times {\rm U(1)}  \, , \\
& \left<78\right> &
\end{eqnarray*}
 but this is not possible with scalars in the 27:
\begin{eqnarray*}
\rm E_6 & \slash{\!\!\!\longrightarrow} & {\rm SU(3)} \times {\rm SU(2)}
\times {\rm U(1)}.\\
& \left<27\right> &
\end{eqnarray*}
 Thus superstrings and {\sc susy} {\sc gut}s appear to be not truly
compatible!
There is unification in superstrings, but it occurs at the
Planck scale, with\cite{RW}
\[
g_3^3 k_3 \ = \ g_2^2 k_2 \ = \ g_2^2 k_1
\ = \ \frac{4\pi G_{\rm N}}{\alpha'},
\]
 where $G_{\rm N}$ is Newton's constant
and $\alpha'$ is the string slope. Here
$k_1$, $k_2$ and $k_3$ are numbers associated with
the Kac-Moody levels of the string ({\it e.g.}
$k_3 = k_2 =1$, $k_1 = 5/3$.)
Thus the `experimental' unification of couplings
at $M_X = 10^{16}\;{\rm GeV}$ appears to be a coincidence
for superstring theory!
What happens in many examples is that the compact manifold $K$
is not simply connected.  In this case then non-trivial
gauge field configurations (Wilson loops) can be trapped
in $K$, leading to a breakdown of the gauge group, much
as if there was an effective scalar in the adjoint\cite{HW}.
Typical breakdown chains for $E_6$ produced by this, so called,
Hosotani-Witten mechanism are \cite{Chains}:
\begin{eqnarray*}
E_6 & \longrightarrow & SU(3) \times SU(2) \times U(1)^n \\
E_6 & \longrightarrow & SU(5) \times U(1) \\
E_6 & \longrightarrow & SU(3)\times SU(3) \times SU(3)
\end{eqnarray*}
To get rid of some of the extra gauge
interactions--as in the case of the
flipped SU(5) chain\cite{Flip}--one needs
some intermediate scale breaking, something that
is also quite against the simple picture of {\sc susy gut}
unification discussed earlier.
\item{ii)} More on the positive side, compactification (either real
or  effective) has a very nice advantage since it easily
generates Yukawa couplings.  The $\rm D>4$ gauge couplings can
be transmuted into Yukawa couplings in four dimensions,
since the gauge fields in the extra dimensions become
Higgs fields in four dimensions after compactification.
This is shown schematically in Fig. 24.
\end{description}
\vbox{
\vspace{7cm}
%\begin{figure*}[h]
%\vspace{6cm}
\begin{center}
{{\bf Figure 24.} \it Generation of
Yukawa couplings through compactification.}
\end{center}}
%\end{figure*}
\noindent
Although one cannot really compute the Yukawa couplings
(that would require a complete normal mode decomposition
of $\Psi$ and $A^a$ on the compact space)
one can, in certain circumstances,
infer some information for a given manifold $K$
on discrete symmetries which can give the overall
texture of the mass matrix (\eg which elements are zero).
Some encouraging results exist\cite{GMR}, but these mass
matrices are not quite ready to be confronted with
experiment yet!

\section*{Concluding Remarks}

I hope I have given you
in these lectures a panorama of the ideas that have been put
forth to understand the deep structures underlying the {\sc sm},
summarised by the 3 questions:
\begin{itemize}
\item[\ ] the question of mass;
\item[\ ] the question of forces;
\item[\ ] the question of matter.
\end{itemize}
 I do not believe, however, that the answers to these questions will come
forth from purely theoretical arguments, although a good theory
argument is always nice!  But I do
believe that there is a good chance that,
with the experiments now being planned with present accelerators
(before the year 2000) and, certainly, with future accelerators
({\sc lhc} and an $e^+e^-$ linear collider---before the year 2010)
we will get a glimmer of which (if any!) of these ideas correctly
describe nature.

\section{Acknowledgements}
I am grateful to David Saxon and Carlos Garcia Carnal for having invited
me to lecture at St. Andrews and at Mar del Plata, respectively.  Both
the Scottish Summer School and the Escuela Latino Americana de Fisica
had an extremely spirited and very engaged group of students who made
the delivery of my lectures a real challenge and a great pleasure.
I also owe a great debt of gratitude to
Lance Vick, for without his invaluable help
the written version of my notes would have remained a chimera.  This
work was supported in part by the Department of Energy under Grant DE-FG03-
91ER40662.

%%%%%%%%%%%%%%%%%%%%%%%%%%
%%%%%%%%%%%%%%% REFERENCES
%%%%%%%%%%%%%%%%%%%%%%%%%%
%\section*{References}
%\frenchspacing
%\begin{small}
%\reference{Achucarro A and Vachaspati T, 1991, {\em Phys Rev D\/}\vol44 3067.}
%\reference{Anderson G and Hall L J, 1992, {\em Phys. Rev. D}\vol45 2685.}
%\reference{Abbott L F and Wise M B, 1984, {\em Nucl Phys B}\vol244 541.}

%\end{small}

\end{document}